\documentclass[11pt]{article}

\usepackage{amsmath,mathrsfs,amsbsy,amsthm}
\usepackage{cite}
\usepackage{graphicx,epsfig}
\usepackage{latexsym,amssymb}
\usepackage{epsf}
\usepackage{ifpdf,lineno}
\usepackage{color}
\usepackage{geometry}
\usepackage{xcolor}

\begin{document} 
	
	\title{{\bf Relativistic $k$-fields with  Massless Soliton Solutions in $3+1$ Dimensions}}
	
	\author{M. Mohammadi$^{1}$\thanks{Corresponding Author. } \\
	{\small \texttt{physmohammadi@pgu.ac.ir}}\and
	R. Gheisari$^{1}$ }
	\date{{\em{$^1$Physics Department, Persian Gulf University, Bushehr 75169, Iran.}}}
	\maketitle
	
\begin{abstract}
 In this work,  the relativistic non-standard  Lagrangian densities  ($k$-fields) with massless solutions are  generally  introduced. Such solutions are not necessarily  energetically   stable. However, in  $3+1$ dimensions, we  introduce a new $k$-field model that results in a single non-topological massless solitary wave solution.   This special solution is energetically stable; that is, any arbitrary deformation above its background leads to an increase in the total energy. In other words, its energy is zero which is the  least energy in all solutions. Hence, it can be called a massless soliton solution. 
\end{abstract}


	

	\section{Introduction}\label{sec1}

The soliton and solitary wave solutions   of the relativistic classical field theory have been a matter of interest in recent decades.  
They behave like  classical   particles and properly satisfy the standard relativistic relations
\cite{rajarama,TS}. Solitary wave solutions or lumps are special traveling wave solutions with localized energy density functions.  
A soliton solution is typically defined as a special stable solitary wave solution that reappears after any collision without any distortion \cite{rajarama}. However, in this paper, we only accept the stability condition for defining soliton solutions. 
Solitary wave solutions are divided into topological and non-topological groups based on their boundary behavior at infinity.
Topological solitary wave solutions  are inevitably stable, and  they are all  solitons. 
The well-known topological kink (anti-kink) solutions of the real nonlinear Klein-Gordon (KG) systems are  good examples of the topological solitons in 1+1 dimensions \cite{rajarama,TS,TA1,TA2,TA3,Kink8}. Also, in $3+1$ dimensions, the    solitons   of the Skyrme model \cite{TS,SKrme,SKrme2,SKrme3,SKrme55} and magnetic monopole  solutions  of the 't Hooft-Polyakov model  \cite{rajarama,TS,toft,pol,TOF} are the well-known topological solutions of the nonlinear relativistic classical  field systems. 
In general, there is extensive literature on topological solitons, as seen in  \cite{CTOPO} and the references therein.

In relation to non-topological soliton solutions, most physical models are non-relativistic. For example, the KdV equation  or the nonlinear Schr\"{o}dinger equation are well-known examples of this type. The nonlinear Schr\"{o}dinger  equation and its different variants have been studied extensively  in nonlinear optics \cite{Optic1,Optic2,Optic3,Optic4,Optic5,Optic6,Optic7,Optic8,Optic9,Optic10,Optic11,Optic12,Optic13,Optic14,Optic15,Optic16},  applied mathematics \cite{Am1,Am2,Am3,Am4} and plasma \cite{Plasma1,Plasma2,Plasma3,Plasma4,Plasma5}. 
Also, the modified versions of the KdV equation have been of great interest to researchers in applied physics and mathematics \cite{KDV1,KDV2,KDV3,KDV4,KDV5}.
The famous relativistic non-topological solitary wave solutions are Q-balls \cite{R2,R8,Vak3,Vak4,Vak5,Vak6}. Since there is no dependence on the boundaries for the non-topological solitary wave solutions,  various criteria have been introduced for the stability considerations.   
Three stability criteria have been introduced especially for the Q-balls: the classical, the quantum mechanical, and the fission criteria \cite{Vak3,Vak4,Vak5,Vak6,Vak22}. 
However, the classical criterion is the most important of all.
It is based on dynamical equations obtained for the small fluctuations above the background of the non-topological solitary wave solutions. 
Above all,  if it can be proved that for any arbitrary deformation in the internal structure of a relativistic  solitary wave solution, the total energy always increases, it would be an energetically stable solution. 
The rest energy would be minimal for such a solution compared to  other (close) solutions; therefore, the solution would be inevitably stable \cite{Derrick,MG,MA,MPS,MPS1}. 
For example, the well-known kink (antikinks) solutions are energetically stable entities 
\cite{rajarama,Derrick}.


All relativistic soliton and solitary wave solutions  that have been introduced so far, have  non-zero rest-masses.   The question  is whether it is possible to have a relativistic  soliton solution with a zero rest-mass.  In general, any  particle  which moves at the speed of light must be massless. But does any massless particle-like entity    have  to move at the speed of light?  In other words,
is it possible to have a zero  rest-mass  particle-like entity   which is at rest or moving  at any arbitrary velocity? Mathematically, if we use  classical relativistic field theory with soliton solutions, our answer may be slightly different.  In  \cite{MG}, it was shown  that the existence of a non-moving massless soliton solution can be possible theoretically in $1+1$ dimensions.
In this paper, we also show that the existence of the relativistic massless  solitons  in $3+1$ dimensions are theoretically possible.


To obtain a  stable zero rest-mass   soliton solution, we need to use a special type of   non-standard 
Lagrangian (NSL) densities for  relativistic fields.  Briefly, for a set of the real scalar fields $\phi_{j}$ ($j=1,2,\cdots,N$),   NSL densities  are not linear in the kinetic scalars, which are also called extended KG systems in \cite{MG,MA,MPS,MPS1}. The kinetic scalars are different contractions  of the scalar fields' derivatives, i.e. ${\cal S}_{ij}=\partial_{\mu}\phi_{i}\partial^{\mu}\phi_{j}$. 
The so-called k-fields, fields with dynamics governed by a non-standard kinetic term, is another name for  such systems \cite{Baz1,Adam,Bab}.  
Historically, the non-standard Lagrangians  were first named  by Arnold \cite{Arnold} and studied for  dynamical systems with a finite number of degrees of freedom, especially in the works of scholars such as Musielak \cite{DY1,DY2,DY22} and El-Nabulsi \cite{DY3,DY4,DY5,DY6,DY7}. In this regard, the non-standard exponential Lagrangians (NSELs),  non-standard power-law Lagrangians (NSPLs), and non-standard Logarithmic Lagrangians (NSLLs) are three types of NSLs that have received more attention in recent studies   \cite{El1,Song,Noether,Geometric0,Geometric,RAm,Rami}.
There is a wide range of applications for such systems concerning in differential equations \cite{Diff1,Diff2,Diff3,Diff4,SA2},  classical and quantum field theory \cite{R2,DY3,qu1,qu2,Cfield,Cfield2,Armend,Chiba,Armend2,cosmology1,Armend3,cos2,cos3,Alekseev},   stellar dynamics \cite{Milky}, and plasma waves \cite{ple}. 
In cosmology, the $k$-field models are particularly popular. They are proposed in inflation theory leading to $k$-inflation \cite{Armend,Chiba,Armend2,cosmology1}, or used to describe dark energy and dark matter \cite{Armend3,R2,cos2,cos3,cos4}.   Classical Yang–Mills field theories with NSL densities can also be  used  in quantum chromodynamics to explain quark-antiquark interactions at large distances \cite{Alekseev}.

The organization of this paper is as follows: in  the next section,  the $k$-field systems with zero rest-mass solutions will be  generally introduced. A preliminary $k$-field model will be introduced to illustrate some aspects of the  stability of a massless  solution.
In section 3,  a new  $k$-field system in $3+1$ dimensions  will be  introduced that  yields  to a single massless energetically stable solitary wave solution.   The last section is devoted to  conclusions.

\section{Massless solutions}\label{sec2}

First of all, let us explain   the conditions that must  be imposed  if we want to have a massless solitary wave solution (defect structure).
For a set of relativistic  scalar  fields $\phi_{k}$ ($k=1,\cdots, N$), the standard Lagrangian densities  are functions of the fields and the kinetic scalers ${\cal S}_{ij}=\partial_{\mu}\phi_{i}\partial^{\mu}\phi_{j}$:
 \begin{equation} \label{lags}
{\cal L}= {\cal L}(\phi_{k},{\cal S}_{ij}),\quad\quad (i,j,k=1,\cdots, N)
 \end{equation}
where $\phi_{k,\mu}=\frac{\partial \phi_{k}}{\partial x^{\mu}}$, and $x^{\mu}\equiv (t,x,y,z)$ \footnote {Note that we set the
	speed of light to one (c = 1) throughout the paper for the sake of simplicity.}. According to the principle of least action, the dynamical  equations of motion would be,
 \begin{equation} \label{lfg}
\frac{\partial{\cal L}}{\partial\phi_{i}}-\frac{\partial}{\partial x^{\mu}}\left(\frac{\partial {\cal L}}{\partial (\partial_{\mu}\phi_{i})}\right)=\frac{\partial{\cal L}}{\partial\phi_{i}}-\sum_{j=1}^{N}\left[\frac{\partial}{\partial x^{\mu}}\left(\frac{\partial {\cal L}}{\partial {\cal S}_{ij}}\right)\partial^{\mu}\phi_{j}+\frac{\partial{\cal L}}{\partial {\cal S}_{ij}}\partial_{\mu}\partial^{\mu}\phi_{j}\right]=0.
 \end{equation}
In general, since Lagrangian density (\ref{lags}) is    invariant under the infinitesimal space-time translations,  four continuity equations $\partial_{\mu}T^{\mu\nu}=0$ and then four conserved quantities $P^{\mu}=\int T^{o\mu}d^{3}\textbf{x}$  are  obtained,   where
 \begin{equation} \label{egb}
T^{\mu\nu}=\sum_{i=1}^{N}\frac{\partial{\cal L}}{\partial\phi_{i,\mu}}\frac{\partial\phi_{i}}{\partial x_{\nu}}-{\cal L}g^{\mu\nu},
\end{equation}
is called the energy-momentum tensor and $g^{\mu\nu}$  is   the $3 + 1$ dimensional Minkowski metric. The energy density function $\varepsilon$ is the   $T^{00}$ component of the energy-momentum tensor (\ref{egb}):
 \begin{equation} \label{e5b}
T^{00}=\varepsilon=\sum_{i=1}^{N}\frac{\partial{\cal L}}{\partial\dot{\phi}_{i}}\dot{\phi}_{i}-{\cal L}=\sum_{j=1}^{N}\sum_{i=1}^{N}\frac{\partial{\cal L}}{\partial{\cal S}_{ij}}\dot{\phi}_{i}\dot{\phi}_{j}(\delta_{ij}+1)-{\cal L}.
\end{equation}
A special localized solution whose  energy density function (\ref{e5b}) is zero everywhere (i.e. $\varepsilon=0$) can be introduced as a zero rest-mass (massless) solitary wave solution. 
 Also, a zero rest-mass solution clearly has to  satisfy  dynamical equations (\ref{lfg}). Thus, condition $\varepsilon=0$ can be assumed  as a new partial differential equation (PDE) along with $N$ coupled PDEs (\ref{lfg}). Naturally,   the existence of $N+1$  coupled  PDEs  for $N$ fields is  scarcely expected  to have a solution. However, if the Lagrangian density  and all its derivatives, i.e. ${\cal L}$, $\frac{\partial{\cal L}}{\partial\phi_{i}}$, $\frac{\partial{\cal L}}{\partial{\cal S}_{ij}}$, and $ \frac{\partial}{\partial x^{\mu}}(\frac{\partial {\cal L}}{\partial {\cal S}_{ij}})$,  become zero for a special  solution, these $N+1$ PDEs  will no doubt be satisfied  automatically and the special solution would be a zero rest-mass solution.


Accordingly, it is easy to understand, based on any standard Lagrangian density $\mathbb{L}$, for which there is   a special solution for condition $\mathbb{L}=0$,  a new Lagrangian density  with a zero rest-mass  solution can be introduced as a power of $\mathbb{L}$, i.e. ${\cal L}=\mathbb{L}^{n}$ provided that $n\geqslant 3$.
For example, for a single scalar field $\phi$ with a standard nonlinear KG Lagrangian density $\mathbb{L}=\partial_{\mu}\phi\partial^{\mu}\phi+4\phi^3-4\phi^4$,   there is a solution for condition $\mathbb{L}=0$, i.e. $\phi=1/(1+x^2+y^2+z^2)$.  This  solution would be a canonical zero rest-mass solution for a Lagrangian density (${\cal L}=\mathbb{L}^3$) as well. In fact, for ${\cal L}=\mathbb{L}^3$ we have $\frac{\partial{\cal L}}{\partial\phi}=3\mathbb{L}^2\frac{\partial\mathbb{L}}{\partial\phi}$, $\frac{\partial{\cal L}}{\partial{\cal S}}=3\mathbb{L}^2\frac{\partial\mathbb{L}}{\partial{\cal S}}$, and $ \frac{\partial}{\partial x^{\mu}}(\frac{\partial {\cal L}}{\partial {\cal S}})=6\mathbb{L}\frac{\partial \mathbb{L}}{\partial x^{\mu}}\frac{\partial \mathbb{L}}{\partial {\cal S}}+3\mathbb{L}^2\frac{\partial}{\partial x^{\mu}}(\frac{\partial \mathbb{L}}{\partial {\cal S}})$, which are obviously  all  zero when $\mathbb{L}=0$.
In general,  for several scalar fields  $\phi_{i}$ ($i=1,\cdots, N$), the Lagrangian density of a  $k$-field system with a zero rest-mass solution will be introduced as follows:     
\begin{eqnarray} \label{sf0}
	{\cal L}=\sum_{n_{1}=0}^{\infty}\sum_{n_{2}=0}^{\infty}\cdots\sum_{n_{m}=0}^{\infty} a({n_{1},\cdots,n_{m}})\mathbb{L}_{1}^{n_{1}}\mathbb{L}_{2}^{n_{2}}\cdots\mathbb{L}_{m}^{n_{m}},
\end{eqnarray}
where $\mathbb{L}_{j}$'s ($j=1,\cdots,m$) are a number of independent Lagrangian densities all of which are zero simultaneously  for the zero rest-mass solution (i.e., $\mathbb{L}_{j}=0$),  provided $n_{1}+n_{2}+\cdots+n_{m} \geqslant 3$.    Note that    coefficients $a({n_{1},\cdots,n_{m}})$  can be  arbitrary  functions  of the fields and the kinetic scalers ${\cal S}_{ij}$. This form of the Lagrangian densities (\ref{sf0}) is very similar to  NSPLs introduced by  El-Nabulsi  for dynamical systems with finite degrees of freedom  \cite{DY7,El1}.

So far, we have only explained how the Lagrangian density of a system of fields must  yield a massless solution, but we have not considered   the stability of such special solutions. 
The energetical stability condition imposes severe constraints on the Lagrangian density (\ref{sf0}), which causes series (\ref{sf0}) to be converted to special formats.
In fact, no rule has been found yet to develop  a system with a single energetically  stable massless  solitary wave solution, and development of such a system would be  mostly based on  trial and error. In this section and the next, we will  try to show  some of  the problems of finding a $k$-field system with a single  energetically  stable zero rest-mass  solution.

According  to the same  $k$-field model in $1+1$ dimensions which was introduced in  \cite{MG} and led  to a single  massless  solitary wave solution, one can think about the modified  version of that in $3+1$ dimensions.
In other words, exactly the same  Lagrangian density which was introduced in $1+1$  dimensions (Eq.~15 in  \cite{MG}) for two scalar fields $\phi_{1}=R$ and  $\phi_{2}=\theta$ is used here again:
 \begin{equation} \label{kexn}
{\cal L}=\sum_{i=1}^3{\cal K}_{i}^3,
 \end{equation}
where
 \begin{eqnarray} \label{e5}
&&{\cal K}_{1}=R^2\mathbb{L}_{2},\\&&\label{e52}
{\cal K}_{2}=R^2\mathbb{L}_{2}+\mathbb{L}_{1}, \\&&\label{e53}
{\cal K}_{3}=R^2\mathbb{L}_{2}+\mathbb{L}_{1}+2R\mathbb{L}_{3},
\end{eqnarray}
 in which, $\mathbb{L}_{2}={\cal S}_{22}-2$, $\mathbb{L}_{1}={\cal S}_{11}-4R^4+4R^3$,  $\mathbb{L}_{3}={\cal S}_{12}$,  ${\cal S}_{11}=\partial_{\mu}R\partial^{\mu}R$, ${\cal S}_{22}=\partial_{\mu}\theta\partial^{\mu}\theta$ and ${\cal S}_{12}=\partial_{\mu}R\partial^{\mu}\theta$.

 Now, the main modification is that the kinetic scalars are defined in the $3 + 1$ dimensions; namely, ${\cal S}_{11}=\partial_{\mu}R\partial^{\mu}R=(\frac{\partial R}{\partial t})^2-(\nabla R)^2$, and so on.   Thus, all the equations of motion and    energy density relations (i.e. equations (19)-(26) in \cite{MG}) would be obtained  again provided one changes  $R'$ and $\theta'$ (i.e. the $x$-derivative of  the module and phase field) to $\nabla R$ and $\nabla \theta$, respectively. In  \cite{MG}, it was shown that the existence of a massless solitary wave solution is possible  if  all ${\cal K}_{i}$'s or $\mathbb{L}_{i}$'s are zero simultaneously. Hence, for ${\cal K}_{i}=0$ ($i=1,2,3$), there was just a unique  non-trivial common  solitary wave  solution as follows:
\begin{equation} \label{So}
R(x)=\dfrac{1}{1+x^2},\quad  \quad\theta(t)=\pm\sqrt{2}t.
\end{equation}
In the $3+1$ dimensions, the required  conditions ${\cal K}_{i}=0$ ($i=1,2,3$)  lead to the following covariant PDE's:
 \begin{eqnarray} \label{dfb}
&&\partial_{\mu}\theta\partial^{\mu}\theta=\dot{\theta}^2-(\nabla \theta)^2=2,\\&& \label{dfb2}
\partial_{\mu}R\partial^{\mu}R=\dot{R}^2-(\nabla R)^2=4R^4-4R^3, \\&& \label{dfb3}
\partial_{\mu}R\partial^{\mu}\theta=\dot{\theta}\dot{R}-(\nabla \theta\cdot \nabla R)=0,
\end{eqnarray}
where, the dot  indicates the time derivative. In general, since there are  three independent PDE's (\ref{dfb})-(\ref{dfb3}) just for two scalar fields $R$ and $\theta$, mathematically, the existence of the common solutions  is  severely restricted. However, for the static massless   solutions for which $\theta(t)=\sqrt{2}t$ and $R=R(x,y,z)$, PDE's (\ref{dfb}) and (\ref{dfb3}) are satisfied  automatically, and PDE (\ref{dfb2}) reduced to
 \begin{equation} \label{hj6}
(\nabla R)^2=\left(\frac{\partial R}{\partial x}\right)^2+\left(\frac{\partial R}{\partial y}\right)^2+\left(\frac{\partial R}{\partial z}\right)^2=4R^3 -4R^4.
\end{equation}
If we restrict  ourselves to  the $1+1$ version of  model (\ref{kexn}) in which $R=R(x)$, the pervious Eq.~(\ref{hj6}) will be reduced   to
 \begin{equation} \label{hj2}
\left(\frac{d R}{dx}\right)^2=4R^3 -4R^4,
\end{equation}
 It is easy to show that   nonlinear ordinary differential  equation~(\ref{hj2}) has just a  unique  non-trivial  solution $R=1/(1+x^2)$, i.e. the one which was introduced in Eq.~(\ref{So}). However, in  the $3+1$  version of the model (\ref{kexn}), the nonlinear  PDE (\ref{hj6}) has infinite solutions, such as the following: 
\begin{eqnarray} \label{jb}
&&R(r)=\dfrac{1}{1+(r+\xi)^2}, \\&&\label{jb2}
R=\frac{1}{1+x^2},\quad R=\frac{1}{1+y^2},\quad R=\frac{1}{1+z^2}\\&&\label{jb3} R=\frac{1}{1+x^2+y^2},\quad  R=\frac{1}{1+x^2+z^2},\quad R=\frac{1}{1+y^2+z^2},
\end{eqnarray}
where $r=\sqrt{x^2+y^2+z^2}$ and $\xi$ is any  arbitrary real number.
According to Eq.~(\ref{jb}), for different  values of $\xi$,  different   degenerate massless  solutions can be obtained in $3+1$ dimensions. In $1 + 1$ version of  model (\ref{kexn}),   static solution (\ref{jb})
 is reduced   to $R = \frac{1}{1+(x+\xi)^2}$, but it is nothing more than a space translation   in  (\ref{So}) and essentially can not be considered as a new special massless solution. Note that special solutions (\ref{jb2}) and (\ref{jb3}) are non-localized  and cannot be physically interesting.

In  \cite{MG} or the  $1+1$ version of  model (\ref{kexn}),  the main point  which  guides one  to conclude  special solitary wave solution (\ref{So}) is a (massless) soliton solution is the fact that  PDE's (\ref{dfb})-(\ref{dfb3}) are entirely  independent. They have just a unique non-trivial  common  solitary wave solution (\ref{So}). Thus, we ensure that Eq.~(\ref{So}) is a single massless  solution with  the minimum energy of all   solutions for the system (\ref{kexn}). In other words, for any arbitrary variation above the background of  single massless solution (\ref{So}), the total energy always increases,    i.e., it is energetically stable and  can be called a soliton solution. 
But, in the $3+1$ version of  model (\ref{kexn}), due to the non-existence of a unique non-trivial common solution for PDEs (\ref{dfb})-(\ref{dfb3}), there is no massless soliton solution. In fact, for PDEs (\ref{dfb})-(\ref{dfb3}),  there is a continuous range of common solutions (\ref{jb})  that are all  degenerate massless solutions of the system.
Hence, they cannot be called  soliton solutions because their profiles can be changed without energy consumption, i.e., there is no stable massless solution.  Accordingly, using two scalar fields $R$ and $\theta$ in the $3+1$ version of  model   (\ref{kexn}) does not lead to a unique  (massless)  common solitary wave  solution for three PDE's (\ref{dfb})-(\ref{dfb3}).  To overcome this problem, the following section will introduce another $k$-field model with three new dynamical fields $\psi_{1}$, $\psi_{2}$ and $\psi_{3}$.


\section{A $k$-field system with a single massless soliton  solution}\label{sec3}

For five real scalar fields $\phi_{1}=R$, $\phi_{2}=\theta$, $\phi_{3}=\psi_{1}$, $\phi_{4}=\psi_{2}$ and $\phi_{5}=\psi_{3}$, we can propose a new $k$-field system in the following form:
 \begin{equation} \label{kk}
{\cal L}=B\sum_{i=1}^{12}{\cal K}_{i}^3,
 \end{equation}
where $B$ can be any arbitrary positive number, and
\begin{eqnarray} \label{jj}
&&{\cal K}_{1}=R^2\mathbb{L}_{2},\quad~~~ {\cal K}_{2}=R^2\mathbb{L}_{2}+\mathbb{L}_{1},\quad~~~ {\cal K}_{3}=R^2\mathbb{L}_{2}+\mathbb{L}_{1}+2 R\mathbb{L}_{3},  \nonumber \\&&\label{jj1}
 {\cal K}_{4}=R^2\mathbb{L}_{2}+\mathbb{L}_{4},\quad~~~ {\cal K}_{5}=R^2\mathbb{L}_{2}+\mathbb{L}_{5},\quad~~~ {\cal K}_{6}=R^2\mathbb{L}_{2}+\mathbb{L}_{6}, \nonumber \\&&\label{jjf}
{\cal K}_{7}=R^2\mathbb{L}_{2}+\mathbb{L}_{4}+\mathbb{L}_{5}+2\mathbb{L}_{7},\quad~~~ {\cal K}_{8}=R^2\mathbb{L}_{2}+\mathbb{L}_{4}+\mathbb{L}_{6}+2\mathbb{L}_{8},\nonumber  \\&&\label{jj6}
{\cal K}_{9}=R^2\mathbb{L}_{2}+\mathbb{L}_{5}+\mathbb{L}_{6}+2\mathbb{L}_{9}, \quad~~~ {\cal K}_{10}=R^2 h_{1} \mathbb{L}_{2}+\mathbb{L}_{1}+\mathbb{L}_{4}+2\mathbb{L}_{10}, \nonumber\\&&\label{jj9}
{\cal K}_{11}=R^2 h_{2} \mathbb{L}_{2}+\mathbb{L}_{1}+\mathbb{L}_{5}+2\mathbb{L}_{11},\quad~~~
{\cal K}_{12}=R^2 h_{3} \mathbb{L}_{2}+\mathbb{L}_{1}+\mathbb{L}_{6}+2\mathbb{L}_{12},
\end{eqnarray}
in which  $h_{j}=[2+\frac{1}{2}(b_{j}-1)^2]$, $b_{j}=2\psi_{j}(2R-1)$ ($j=1,2,3$), and
\begin{eqnarray} \label{ll}
&&\mathbb{L}_{1}={\cal S}_{11}-4R^4+4R^3,\quad~~~ \mathbb{L}_{2}={\cal S}_{22}-2,\quad~~~ \mathbb{L}_{3}={\cal S}_{12},\quad~~~ \mathbb{L}_{4}={\cal S}_{33}+R^2-4R^2 \psi_{1}^2, \nonumber \\&&\label{ll1}
\mathbb{L}_{5}={\cal S}_{44}+R^2-4R^2 \psi_{2}^2,\quad~~~ \mathbb{L}_{6}={\cal S}_{55}+R^2-4R^2 \psi_{3}^2,\quad~~~\mathbb{L}_{7}={\cal S}_{34}-4R^2\psi_{1}\psi_{2}, \nonumber \\&&\label{ll2}
\mathbb{L}_{8}={\cal S}_{35}-4R^2\psi_{1}\psi_{3},\quad~~~
\mathbb{L}_{9}={\cal S}_{45}-4R^2\psi_{2}\psi_{3},\quad~~~  \mathbb{L}_{10}={\cal S}_{13}-b_{1}R^2, \nonumber\\&&\label{ll3}
\mathbb{L}_{11}={\cal S}_{14}-b_{2}R^2,\quad~~~
\mathbb{L}_{12}={\cal S}_{15}-b_{3}R^2,
\end{eqnarray}
where ${\cal S}_{11}=\partial_{\mu}R\partial^{\mu}R$, ${\cal S}_{22}=\partial_{\mu}\theta\partial^{\mu}\theta$, ${\cal S}_{12}=\partial_{\mu}R\partial^{\mu}\theta$, ${\cal S}_{33}=\partial_{\mu}\psi_{1}\partial^{\mu}\psi_{1}$, ${\cal S}_{44}=\partial_{\mu}\psi_{2}\partial^{\mu}\psi_{2}$, ${\cal S}_{55}=\partial_{\mu}\psi_{3}\partial^{\mu}\psi_{3}$, ${\cal S}_{13}=\partial_{\mu}R\partial^{\mu}\psi_{1}$, ${\cal S}_{14}=\partial_{\mu}R\partial^{\mu}\psi_{2}$, ${\cal S}_{15}=\partial_{\mu}R\partial^{\mu}\psi_{3}$, ${\cal S}_{34}=\partial_{\mu}\psi_{1}\partial^{\mu}\psi_{2}$, ${\cal S}_{35}=\partial_{\mu}\psi_{1}\partial^{\mu}\psi_{3}$ and ${\cal S}_{45}=\partial_{\mu}\psi_{2}\partial^{\mu}\psi_{3}$ are  some  kinetic scalars which are used to introduce  the new $k$-field model (\ref{kk}).



Using the  Euler-Lagrange equations, one can easily obtain the following dynamical equations:
\begin{eqnarray} \label{jkt}
&&\sum_{i=1}^{12} {\cal K}_{i}\left[2(\partial_{\mu}{\cal K}_{i})   \frac{\partial{\cal K}_{i}}{\partial(\partial_{\mu}R)}    +   {\cal K}_{i}\partial_{\mu}\left(\frac{\partial{\cal K}_{i}}{\partial(\partial_{\mu}R)}\right)    -    {\cal K}_{i}\frac{\partial{\cal K}_{i}}{\partial R}   \right]=0,\\&&\label{jkt1}
\sum_{i=1}^{12} {\cal K}_{i}\left[2(\partial_{\mu}{\cal K}_{i})   \frac{\partial{\cal K}_{i}}{\partial(\partial_{\mu}\theta)}    +   {\cal K}_{i}\partial_{\mu}\left(\frac{\partial{\cal K}_{i}}{\partial(\partial_{\mu}\theta)}\right)       \right]=0.\\&&\label{jkt2}
\sum_{i=1}^{12} {\cal K}_{i}\left[2(\partial_{\mu}{\cal K}_{i})   \frac{\partial{\cal K}_{i}}{\partial(\partial_{\mu}\psi_{j})}    +   {\cal K}_{i}\partial_{\mu}\left(\frac{\partial{\cal K}_{i}}{\partial(\partial_{\mu}\psi_{j})}\right)    -    {\cal K}_{i}\frac{\partial{\cal K}_{i}}{\partial \psi_{j}}   \right]=0, \quad (j=1,2,3).
\end{eqnarray}
The sets of functions $R$, $\theta$, and $\psi_{j}$ ($j=1,2,3$) for which ${\cal K}_{i}=0$ ($i=1,\cdots,12$)  are simultaneously the   special massless solutions of the new $k$-field model (\ref{kk}). Since   ${\cal K}_{i}$'s are twelve independent linear combinations of twelve  independent scalars   $\mathbb{L}_{i}$'s, it is easy to understand that the conditions ${\cal K}_{i}=0$  are equivalent  to   $\mathbb{L}_{i}=0$ ($i=1,\cdots,12$).  The energy-density  belonging to the new  Lagrangian-density (\ref{kexn})   would be
\begin{eqnarray} \label{nnmn}
&&\varepsilon(x,t)=T^{00}=\sum_{i=1}^{12}\varepsilon_{i}=B\sum_{i=1}^{12}{\cal K}_{i}^{2}\left[3C_{i}
-{\cal K}_{i}\right],
\end{eqnarray}
which are divided  into twelve  distinct  parts, in which
\begin{equation}\label{cof}
C_{i}=\dfrac{\partial{\cal K}_{i}}{\partial \dot{\theta}}\dot{\theta}+\dfrac{\partial{\cal K}_{i}}{\partial \dot{R}}\dot{R}+\sum_{j=1}^{3}\dfrac{\partial{\cal K}_{i}}{\partial \dot{\psi_{j}}}\dot{\psi_{j}}=
\begin{cases}
 2R^2\dot{\theta}^{2} & \text{i=1}
\\
2(\dot{R}^{2}+R^2\dot{\theta}^2) & \text{i=2}
\\2(\dot{R}+R\dot{\theta})^2
 & \text{i=3}.
 \\2(\dot{\psi_{1}}^2+R^2\dot{\theta}^2)
 & \text{i=4}.
 \\2(\dot{\psi_{2}}^2+R^2\dot{\theta}^2)
 & \text{i=5}.
 \\2(\dot{\psi_{3}}^2+R^2\dot{\theta}^2)
 & \text{i=6}.
 \\2(\dot{\psi_{1}}+\dot{\psi_{2}})^2+2R^2\dot{\theta}^2
 & \text{i=7}.
 \\2(\dot{\psi_{1}}+\dot{\psi_{3}})^2+2R^2\dot{\theta}^2
 & \text{i=8}.
 \\2(\dot{\psi_{2}}+\dot{\psi_{3}})^2+2R^2\dot{\theta}^2
 & \text{i=9}.
\\2(\dot{R}+\dot{\psi_{1}})^2+2h_{1}R^2\dot{\theta}^2
 & \text{i=10}.
 \\2(\dot{R}+\dot{\psi_{2}})^2+2h_{2}R^2\dot{\theta}^2
 & \text{i=11}.
 \\2(\dot{R}+\dot{\psi_{3}})^2+2h_{3}R^2\dot{\theta}^2
 & \text{i=12}.
\end{cases}
\end{equation}
After a straightforward calculation we get:
 \begin{eqnarray} \label{eis1}
&&\varepsilon_{1}=B{\cal K}_{1}^2[5R^2\dot{\theta}^2+R^2(\nabla\theta)^2+2R^2],\\ \label{eis2}&&
\varepsilon_{2}=B{\cal K}_{2}^2[5R^2\dot{\theta}^2+5\dot{R}^2+R^2(\nabla\theta)^2+(\nabla R)^2+U(R)],\\ \label{eis3}&&
\varepsilon_{3}=B{\cal K}_{3}^2[5(R\dot{\theta}+\dot{R})^2+(R\nabla\theta+\nabla R)^2+U(R)],\\ \label{eis4}&&
\varepsilon_{4}=B{\cal K}_{4}^2[5R^2\dot{\theta}^2 +5\dot{\psi_{1}}^2+R^2(\nabla \theta)^2+(\nabla\psi_{1})^2+R^2+4R^2\psi_{1}^2],\\
\label{eis5}&&
\varepsilon_{5}=B{\cal K}_{4}^2[5R^2\dot{\theta}^2 +5\dot{\psi_{2}}^2+R^2(\nabla \theta)^2+(\nabla\psi_{2})^2+R^2+4R^2\psi_{2}^2],\\
\label{eis6}&&
\varepsilon_{6}=B{\cal K}_{6}^2[5R^2\dot{\theta}^2 +5\dot{\psi_{3}}^2+R^2(\nabla \theta)^2+(\nabla\psi_{3})^2+R^2+4R^2\psi_{3}^2],\\ \label{eis7}&&
\varepsilon_{7}=B{\cal K}_{7}^2[5R^2\dot{\theta}^2 +5(\dot{\psi_{1}}+\dot{\psi_{2}})^2+R^2(\nabla \theta)^2+(\nabla\psi_{1}+\nabla\psi_{2})^2+4R^2(\psi_{1}+\psi_{2})^2],\\ \label{eis8}&&
\varepsilon_{8}=B{\cal K}_{8}^2[5R^2\dot{\theta}^2 +5(\dot{\psi_{1}}+\dot{\psi_{3}})^2+R^2(\nabla \theta)^2+(\nabla\psi_{1}+\nabla\psi_{3})^2+4R^2(\psi_{1}+\psi_{3})^2],\\ \label{eis9}&&
\varepsilon_{9}=B{\cal K}_{9}^2[5R^2\dot{\theta}^2 +5(\dot{\psi_{2}}+\dot{\psi_{3}})^2+R^2(\nabla \theta)^2+(\nabla\psi_{2}+\nabla\psi_{3})^2+4R^2(\psi_{2}+\psi_{3})^2],\\ \label{eis10}
&& \varepsilon_{10}=B{\cal K}_{10}^2[5h_{1}R^2\dot{\theta}^2+h_{1}R^2(\nabla\theta)^2+5(\dot{R}+\dot{\psi_{1}})^2+(\nabla R+\nabla\psi_{1})^2+V(R,\psi_{1})],\\ \label{eis11}&&
\varepsilon_{11}=B{\cal K}_{11}^2[5h_{2}R^2\dot{\theta}^2+h_{2}R^2(\nabla\theta)^2+5(\dot{R}+\dot{\psi_{2}})^2+(\nabla R+\nabla\psi_{2})^2+V(R,\psi_{2})],
\\ \label{eis12}&&
\varepsilon_{12}=B{\cal K}_{12}^2[5h_{3}R^2\dot{\theta}^2+h_{3}R^2(\nabla\theta)^2+5(\dot{R}+\dot{\psi_{3}})^2+(\nabla R+\nabla\psi_{3})^2+V(R,\psi_{3})],\quad\quad\quad
\end{eqnarray}
where
\begin{eqnarray} \label{UR}
U(R)=4R^4-4R^3+2R^2,
\end{eqnarray}
and
\begin{eqnarray} \label{VR}
V(R,\psi_{j})=U(R)+2R^2+R^2b_{j}^2+4R^2\psi_{j}^2,\quad (j=1,2,3).
\end{eqnarray}
 Both $U(R)$ and $V(R,\psi_{j})$ are  positive definite   functions and  bounded from below by zero. Thus, all terms in
Eqs.~(\ref{eis1})-(\ref{eis6})  are positive definites and  the energy density function (\ref{nnmn}) is also
bounded from below by zero.

\begin{figure}[htp]

  \centering

 \label{pphi1}

  \begin{tabular}{cc}

    \includegraphics[width=28mm]{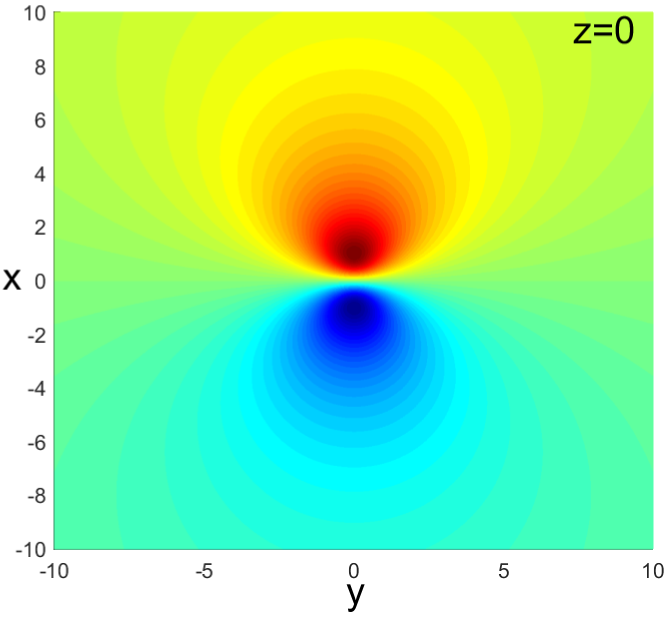}

    \includegraphics[width=28mm]{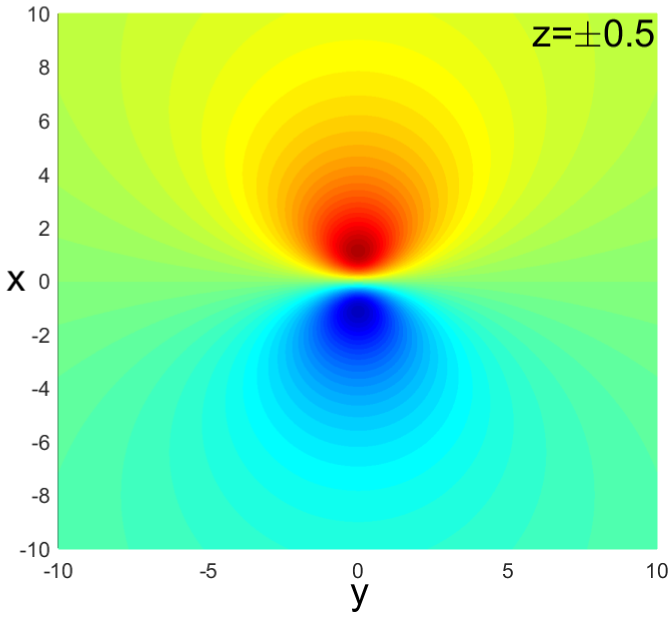}

    \includegraphics[width=28mm]{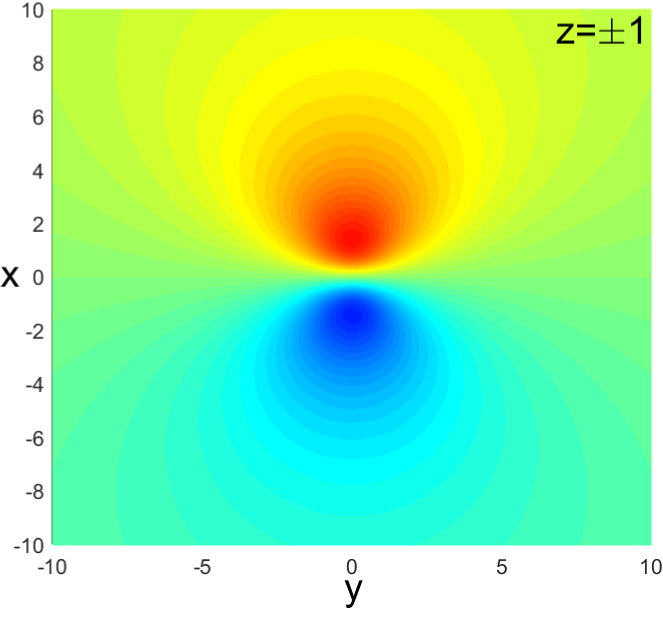}

    \includegraphics[width=28mm]{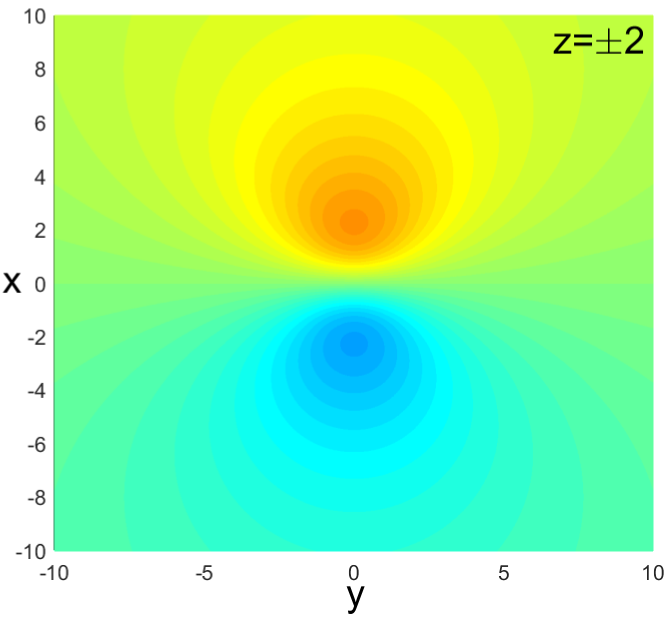}

    \includegraphics[width=33mm]{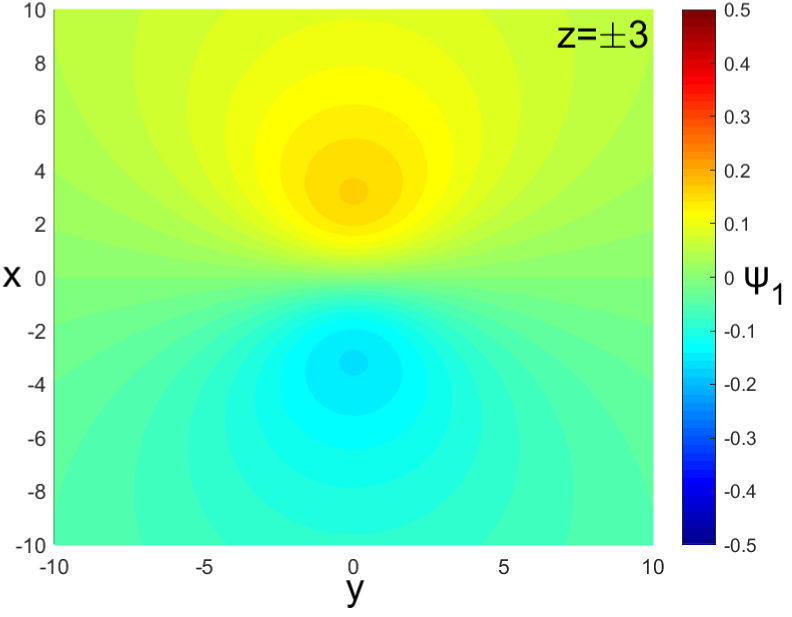}\\

        \includegraphics[width=28mm]{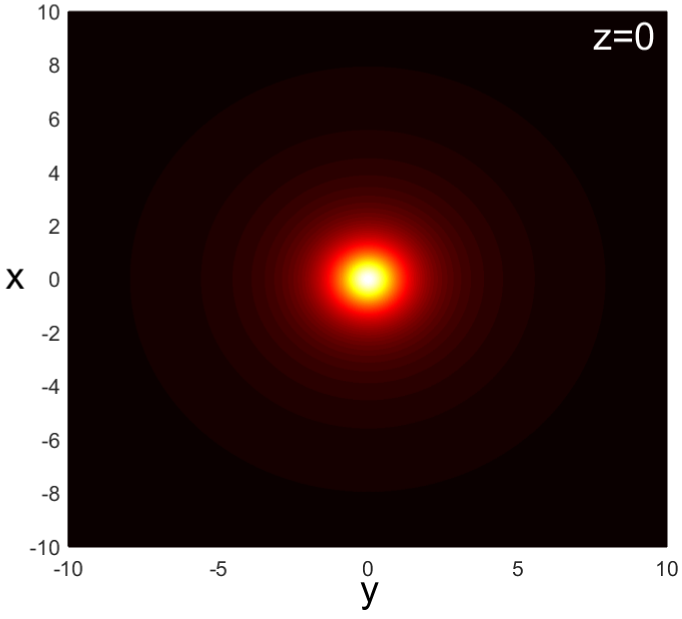}

    \includegraphics[width=28mm]{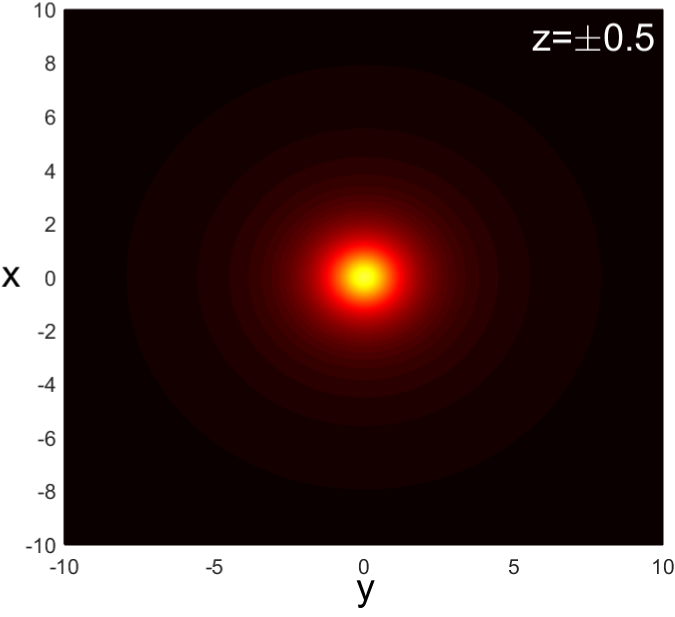}

    \includegraphics[width=28mm]{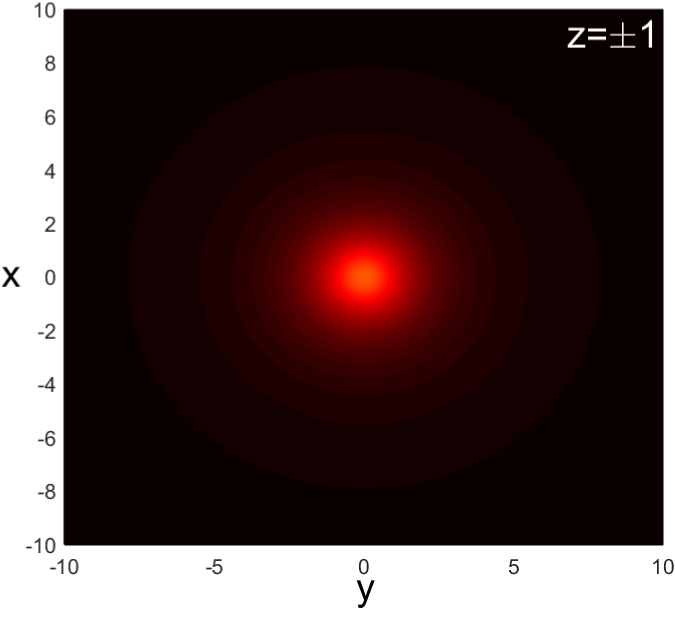}

    \includegraphics[width=28mm]{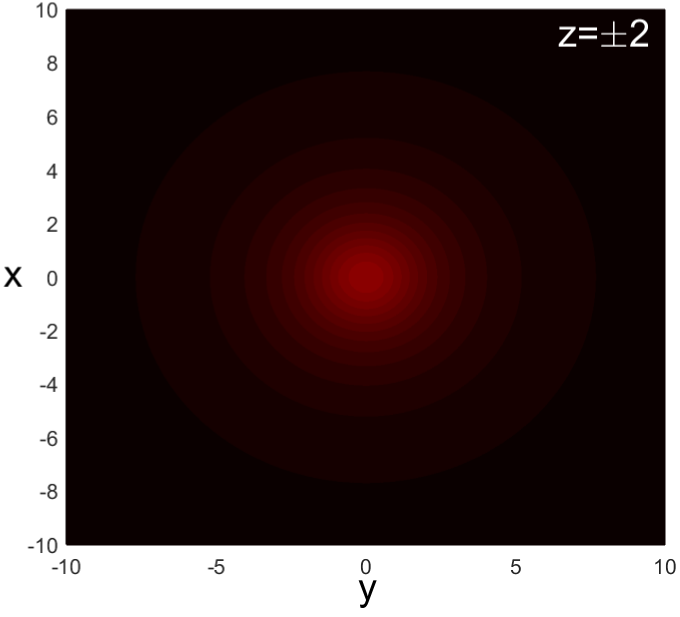}

    \includegraphics[width=33mm]{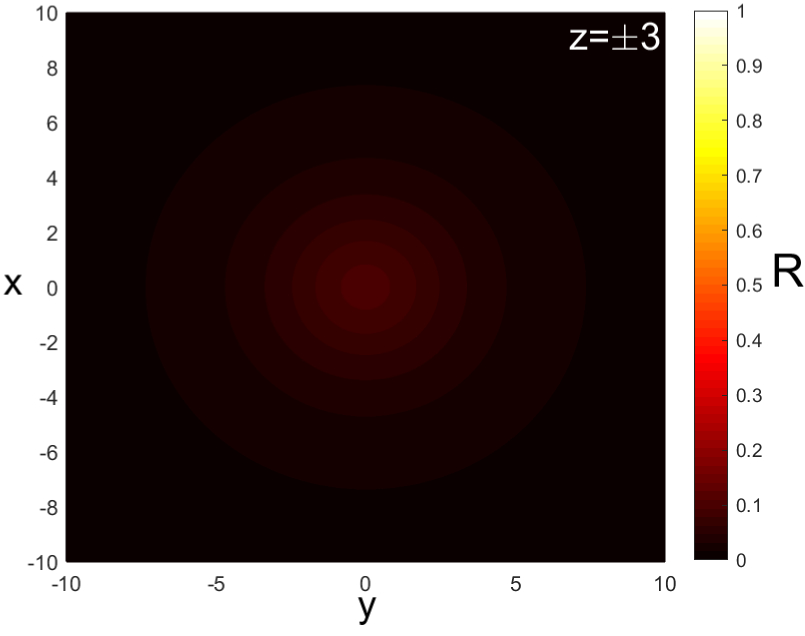}

  \end{tabular}
\caption{The first (second) row  is a four dimensional  scheme for visualizing  $\psi_{1}=x[1+r^2]^{-1}$ ($R=[1+r^2]^{-1}$).}\label{vbg}
\end{figure}

As noted before, a special zero rest-mass solution    would  be possible if $\mathbb{L}_{i}$'s (or equivalently  ${\cal K}_{i}$'s) are zero simultaneously. But,  mathematically,   since there are twelve independent conditions of  $\mathbb{L}_{i}=0$ as  twelve  independent coupled PDE's for just five scalar fields of $R$, $\theta$ and $\psi_{j}$ ($j=1,2,3$),  we normally  do  not expect them  to be satisfied  simultaneously.  However, we build the new $k$-field system (\ref{kk})  in such a way that there is  exceptionally one  massless solution  for which $\mathbb{L}_{i}=0$ as follows:
\begin{eqnarray} \label{SxS}
R=\frac{1}{1+r^2},  \quad \theta=\pm\sqrt{2}t, \quad \psi_{j}=\pm\frac{x^{j}}{1+r^2},\quad (j=1,2,3),
\end{eqnarray}
where $x^{1}=x$, $x^{2}=y$ and $x^{3}=z$ (see Fig.~\ref{vbg}).
Now, unlike the previous  model (\ref{kexn}) with the undesirable degenerate solutions (\ref{jb}), the following set ($\xi\neq 0$) would not be   a special massless solution of the new system (\ref{kk}) anymore:
\begin{eqnarray} \label{As}
R=\frac{1}{1+(r+\xi)^2}, \quad \theta=\pm\sqrt{2}t, \quad \psi_{j}=\pm\frac{x^{j}}{1+r^2}, \quad (\xi\neq 0).
\end{eqnarray}
Moreover,  one can simply check whether or not  the following sets of functions $R$, $\theta$, and $\psi_{j}$ ($j=1,2,3$) are  also  the special solutions of the new system (\ref{kk}); that is, they are not the common solutions  of the PDE's $\mathbb{L}_{i}=0$ ($i=1,\cdots,12$) simultaneously:
\begin{eqnarray} \label{As1}
&& R=\frac{1}{1+(r+\xi)^2}, \quad \theta=\pm\sqrt{2}t, \quad \psi_{j}=\pm\frac{x^{j}}{1+(r+\xi)^2},\quad (\xi\neq 0),\\ \label{As3} &&
R=0, \quad \theta=\pm\sqrt{2}t, \quad \psi_{j}=\pm\frac{x^{j}}{1+r^2},\\ \label{As4}&&
R=\frac{1}{1+r^2}, \quad \theta=\pm\sqrt{2}t, \quad \psi_{j}=0,\\ \label{As5} &&
R=\frac{1}{1+x^2+y^2}, \quad \theta=\pm\sqrt{2}t, \quad \psi_{j}=\pm\frac{x^{j}}{1+x^2+y^2},\\ \label{As6} &&
R=\frac{1}{1+x^2}, \quad \theta=\pm\sqrt{2}t, \quad \psi_{j}=\pm\frac{x^{j}}{1+x^2}.
\end{eqnarray}
It should be noted that  the new system (\ref{kk}) does  not even yield non-localized massless solutions such as Eqs.~(\ref{As5}) and (\ref{As6}). The lack  of  non-localized massless solutions  was the main  reason why we had to use  new scalar fields $\psi_{1}$, $\psi_{2}$, and $\psi_{3}$     to introduce the new system (\ref{kk}).
In fact, we did not succeed in finding a simpler  system with  one or two new scaler fields of $\psi_{1}$ and $\psi_{2}$ without non-localized massless solutions.

In general,  as in the  previous  model (\ref{kexn}), conditions $\mathbb{L}_{2}=\mathbb{L}_{3}=0$ are satisfied  simultaneously for the static solutions  (i.e. $R=R(x,y,z)$,  $\psi_{j}=\psi_{j}(x,y,z)$, and $\theta=\sqrt{2}t$). The static module function $R(x,y,z)$, however, must participate in $10$ completely different PDE's as follows:
\begin{eqnarray} \label{pg}
&&(\nabla R)^2=4R^3-4R^4,\\&&\label{pg2}
\nabla\psi_{i}\cdot\nabla\psi_{j}=R^2\delta_{ij}-4R^2\psi_{i}\psi_{j}\quad (i,j=1,2,3),\\&&\label{pg3}
\nabla\psi_{j}\cdot\nabla R=-2\psi_{j}(2R-1)R^2, \quad (j=1,2,3).
\end{eqnarray}
  Since there are ten  independent PDE's (\ref{pg})-(\ref{pg3}) for four static  scalar fields $R(x,y,z)$ and $\psi_{j}(x,y,z)$  ($j=1,2,3$), mathematically, the possibility of having a common solution is exceptionally low. 
More generally, if we do not restrict ourselves to static solutions,  
 there are twelve independent conditions  $\mathbb{L}_{i}=0$  only for five real scalar fields $R$, $\theta$, and $\psi_{j}$ ($j=1,2,3$). Thus, it is  mathematically very rare to have a common (static or dynamic) solution.
In fact, these coupled equations are built deliberately in such a way to  make Eq.~(\ref{SxS}) an exceptional  static  common solution. In other words,  we first consider Eq.~(\ref{SxS}) and then try to find the proper restrictive conditions $\mathbb{L}_{i}=0$ ($i=1,\cdots,12$) to support it as an outstanding   solution. 
In sum,  it seems  that  the special solution  (\ref{SxS}) is a single  massless solution, and we use this name  in  the rest of the paper. Suppose one succeeds  in finding another massless solution along with (\ref{SxS}). In that case, it would be possible to introduce more complicated systems by imposing  new scalar fields with   additional restrictive conditions  $\mathbb{L}_{i}=0$ to ensure the  uniqueness  of a  massless solitary wave solution.




Since the Lagrangian density (\ref{kk}) is essentially Poincar\'{e} invariant,   any  rotation of  non-spherical symmetric  functions $\psi_{j}=\pm x^{j}(1+r^2)^{-1}$ ($j=1,2,3$)  can be used equivalently in Eq.~(\ref{SxS}).
For example, instead of $\psi_{j}=\pm x^{j} (1+r^2)^{-1}$ ($j=1,2,3$) in Eq.~(\ref{SxS}),  we can use $\psi_{1}=\pm (\cos(\alpha)x+\sin(\alpha)y) (1+r^2)^{-1}$, $\psi_{2}=\pm (-\sin(\alpha)x+\cos(\alpha)y)(1+r^2)^{-1}$ and   $\psi_{3}=\pm z(1+r^2)^{-1}$ (i.e. any arbitrary rotation about $z$-axis), where $\alpha$ is any arbitrary angle. However, since all  different spatial rotations    are  physically equivalent, we can just consider the same simple  functions $\psi_{j}=\pm x^{j} (1+r^2)^{-1}$ ($j=1,2,3$) as the proper   candidates  for  all of them.

According to Eqs.~(\ref{eis1})-(\ref{eis12}), since all terms in   energy density functional (\ref{nnmn}) are positive definites,   this property imposes  a strong  condition to  ensure that  the single massless solution (\ref{SxS}) is really an  energetically  stable object or a soliton solution, which means that  any arbitrary deformation above the background of that leads to an increase in the total energy. Any arbitrary small  deformed version  of  special solution (\ref{SxS}) can be introduced as follows:
\begin{eqnarray} \label{ty}
R=(1+r^2)^{-1}+\delta R, \quad \theta=\sqrt{2}t+\delta\theta,\quad \psi_{j}=\pm x^j(1+r^2)^{-1}+\delta\psi_{j}, \quad (j=1,2,3)
\end{eqnarray}
where $\delta R$, $\delta \theta$, and $\delta \psi_{j}$  (small variations) are  considered to be any arbitrary  small functions  of space-time. If we insert (\ref{ty}) into  $\varepsilon_{i}$ ($i=1,\cdots,12$),  we find
\begin{eqnarray} \label{so4}
&&\delta\varepsilon_{i}=B[3(C_{i}+\delta C_{i})({\cal K}_{i}+\delta{\cal K}_{i})^{2}-({\cal K}_{i}+\delta{\cal K}_{i})^{3}]=B[3(C_{i}+\delta C_{i})(\delta{\cal K}_{i})^{2}-(\delta{\cal K}_{i})^{3}]\approx\nonumber\\&& \quad\quad B[3C_{i}(\delta{\cal K}_{i})^{2}-(\delta{\cal K}_{i})^{3}]\approx[3BC_{i}(\delta{\cal K}_{i})^{2}]>0.\quad\quad
\end{eqnarray}
Note that, for  massless  solution (\ref{SxS}), ${\cal K}_{i}=0$ and $\varepsilon_{i}=0$ ($i=1,\cdots,12$). Hence, since $C_{i}>0$, according to Eq.~(\ref{so4}),
$\delta\varepsilon_{i}$ ($i=1,\cdots,12$) and  $\delta E=\int\sum_{i=1}^{12} \delta\varepsilon_{i} d^3x$ have always positive definite values for all small variations; that is,  massless  solution (\ref{SxS}) is energetically stable.
In other words,
for any arbitrary  deformation above the background of  special solution (\ref{SxS}), at least one of the  ${\cal K}_{i}$'s (or equivalently  one of the  $\varepsilon_{i}$'s) would  be a non-zero functional.  It  leads to the non-zero positive energy density functionals (\ref{so4}),  and then the total energy variation $\delta E$ will be always larger than zero. 
Since  special massless solution (\ref{SxS}) is single, other solutions of the dynamical equations (\ref{jkt})-(\ref{jkt2}) can be considered as the small or large deformations of that. Hence,  they all have non-zero positive rest energies; that is to say,  special solution  (\ref{SxS}) has always the minimum energy  of all.
To summarize, based on the very strict conditions that have been set,  special massless solution (\ref{SxS}) is single and  stable against any arbitrary deformation. Hence, there is no possibility that this particle-like entity (\ref{SxS}) with zero energy will decay  and turn into radiations.

\begin{figure}[htp]

  \centering

  \begin{tabular}{cc}

    \includegraphics[width=36mm]{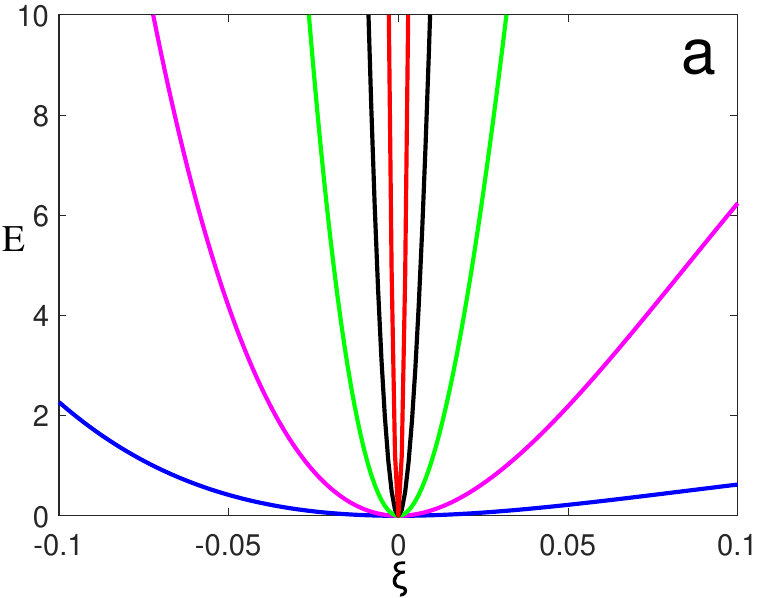}

    \includegraphics[width=36mm]{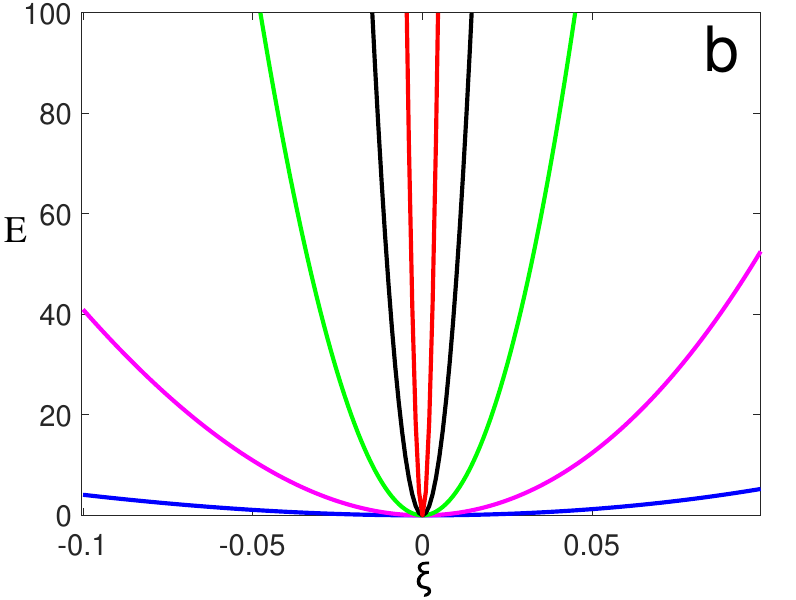}

   \includegraphics[width=36mm]{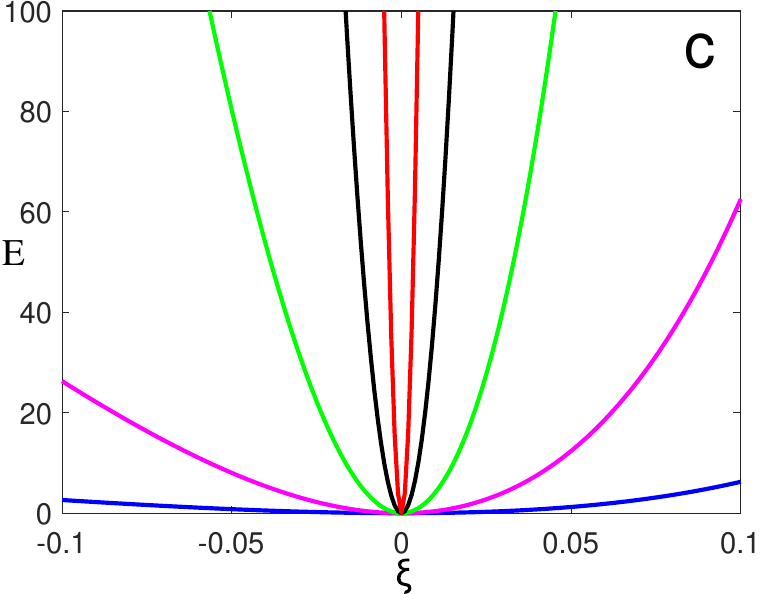}

    \includegraphics[width=36mm]{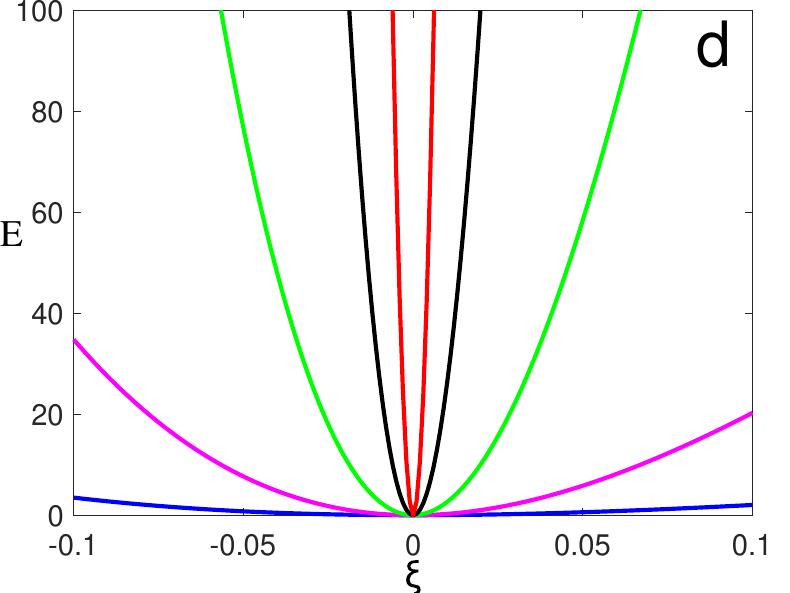}\\

    \includegraphics[width=36mm]{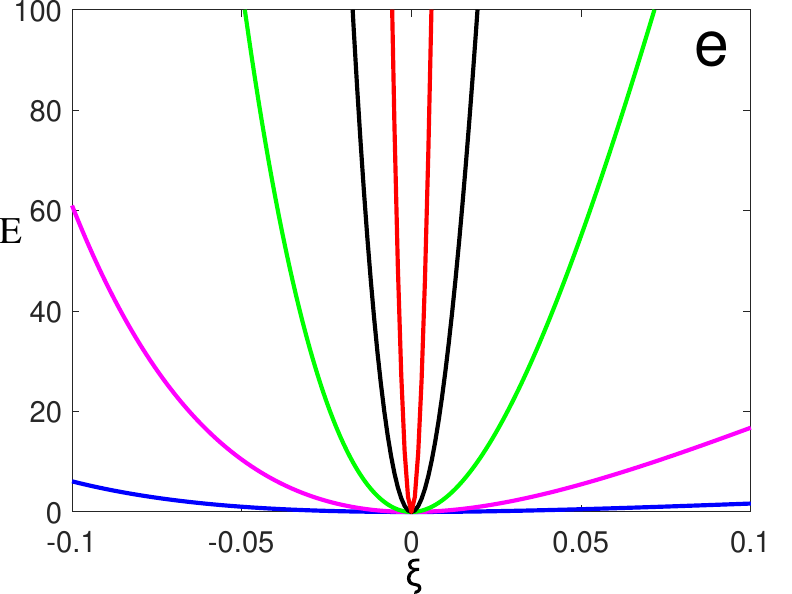}

    \includegraphics[width=36mm]{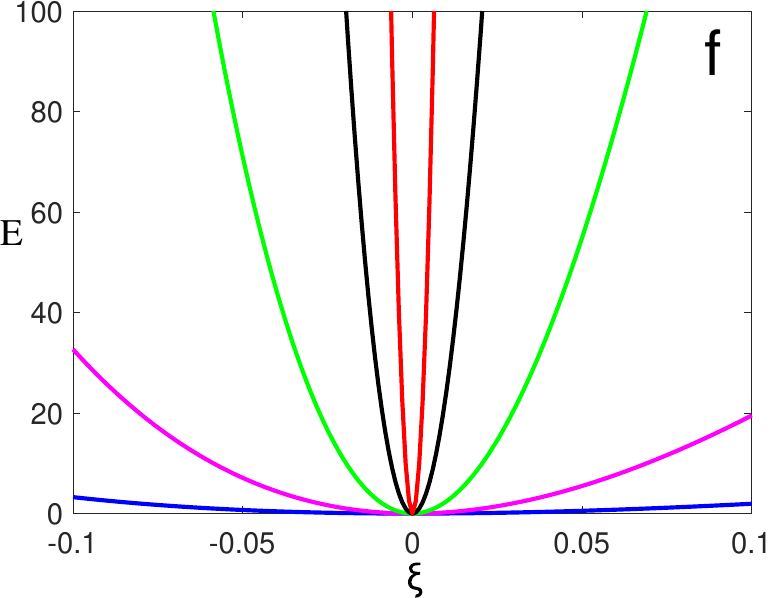}

        \includegraphics[width=36mm]{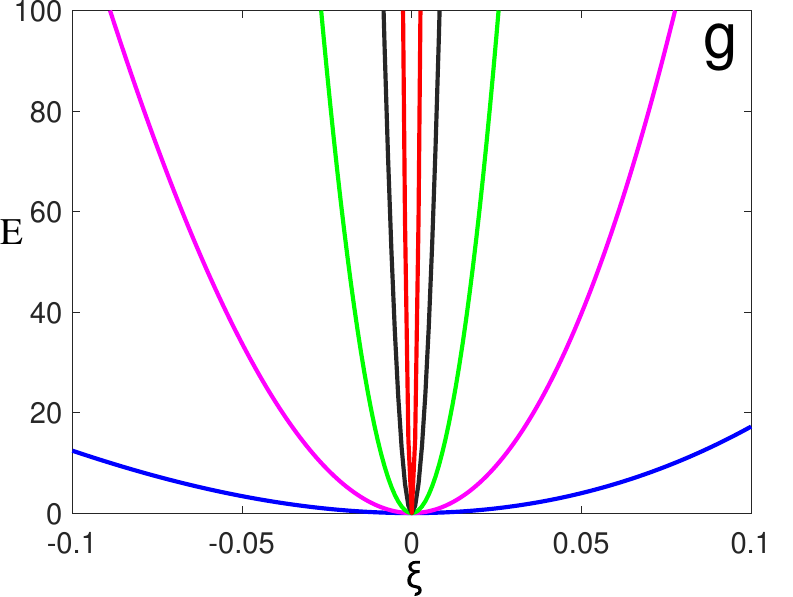}

    \includegraphics[width=36mm]{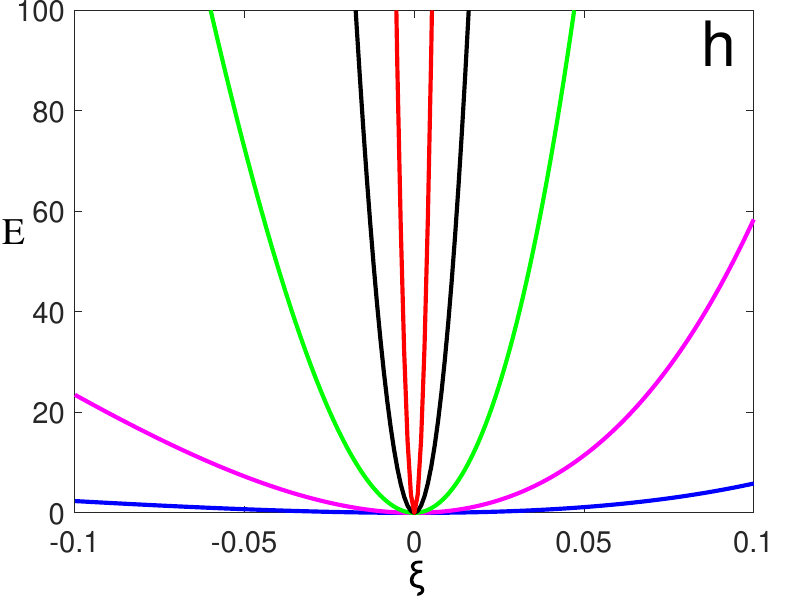}\\

    \includegraphics[width=36mm]{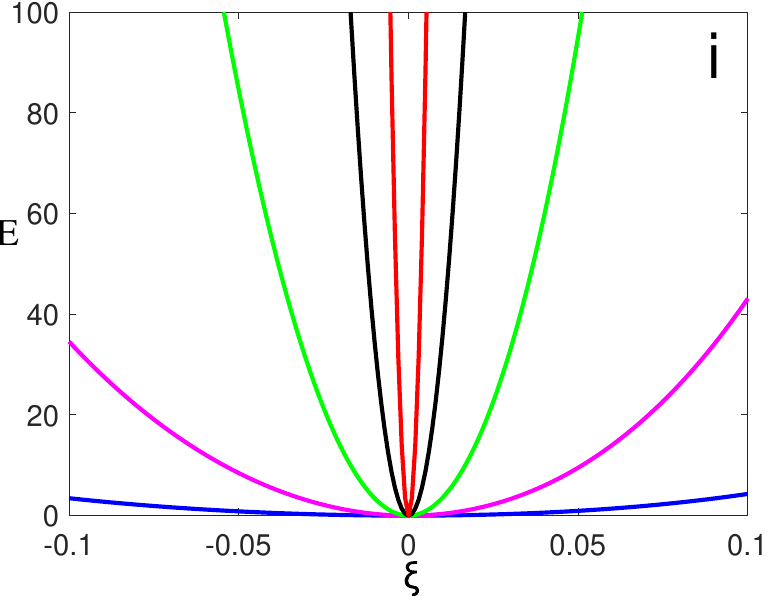}

    \includegraphics[width=36mm]{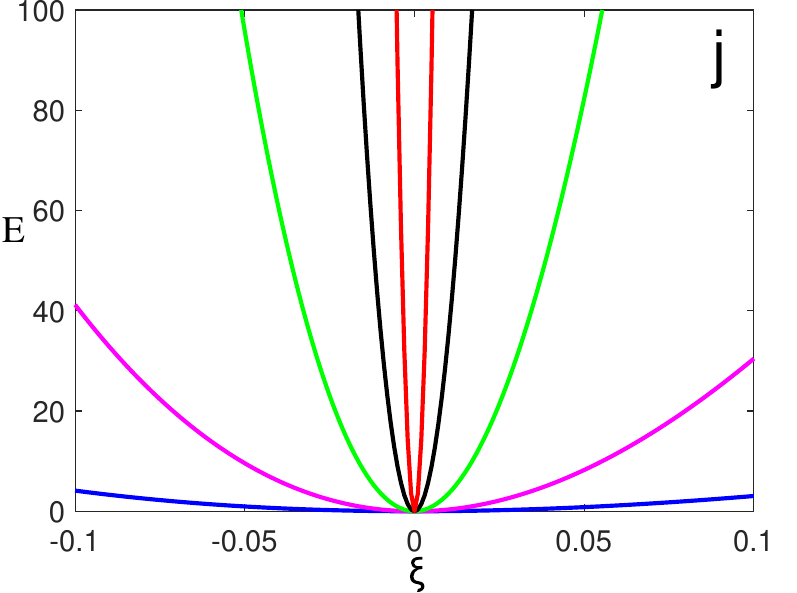}

    \includegraphics[width=36mm]{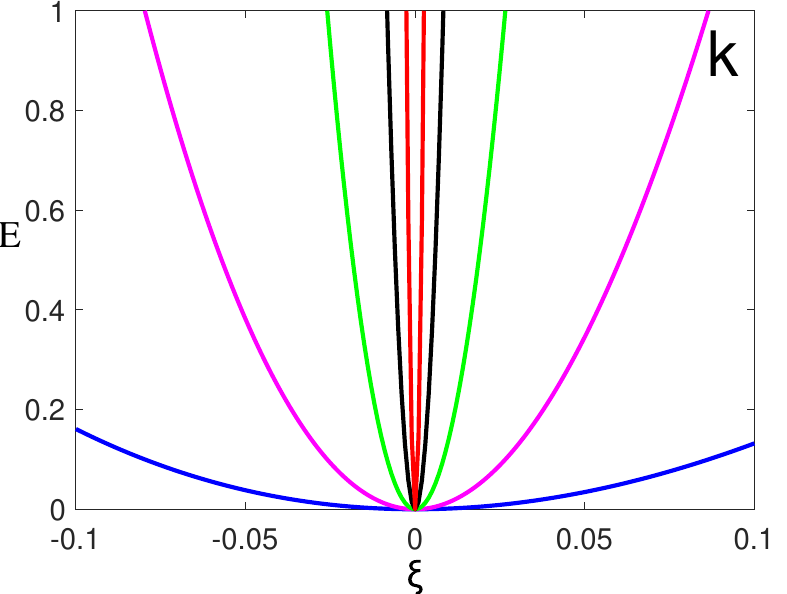}

    \includegraphics[width=36mm]{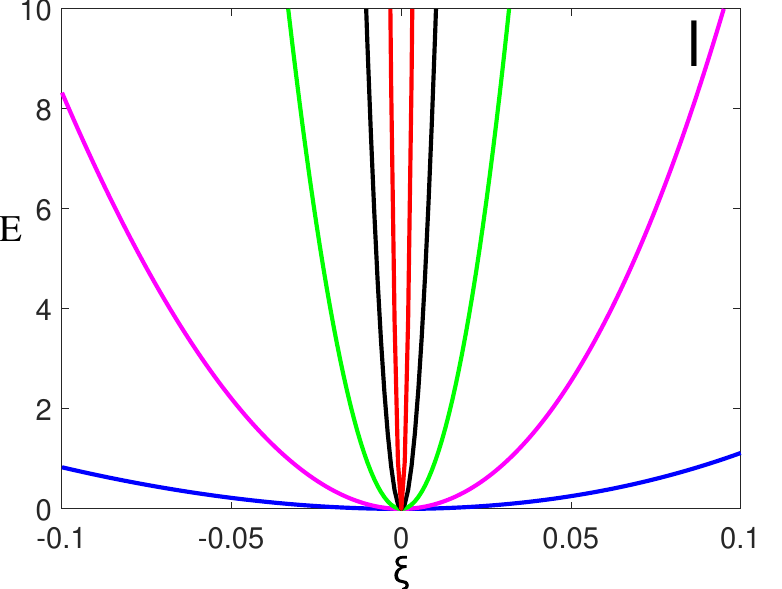}\\

  \end{tabular}
\caption{Plots~a-l represent  variations of the total energy $E$ versus small $\xi$ for different deformations (\ref{As1}) and (\ref{var1})-(\ref{var10})  at $t = 0$, respectively. Various
color curves of blue, purple, green, black, and red are related to $B=1$, $B=10$, $B=100$, $B=1000$, and $B=10000$, respectively.}\label{mnb}
\end{figure}

For more support, let us consider  the total energy variation ($E=\delta E$) for many arbitrary small deformations above the background of  special massless solution (\ref{SxS}) numerically. For example,  a number of arbitrary ad hoc   deformations can be the same as the one  introduced in Eq.~(\ref{As1}) and  eleven  other cases as follows:
\begin{eqnarray} \label{var1}
&&R=(1+\xi)(1+r^2)^{-1}, \quad \theta=\pm\sqrt{2}t, \quad \psi_{j}=\pm x^{j}(1+r^2)^{-1},\\  \label{As2} &&
R=(1+r^2)^{-1}, \quad \theta=\pm\sqrt{2}t, \quad \psi_{j}=\pm(1+\xi)x^{j}(1+r^2)^{-1},\\&&\label{var2}
R=(1+(r+\xi)^2)^{-1}, \quad \theta=\pm\sqrt{2}t, \quad \psi_{j}=\pm x^{j}(1+r^2)^{-1}, \\&&\label{var3}
R=(1+\xi+r^2)^{-1}, \quad \theta=\pm\sqrt{2}t, \quad \psi_{j}=\pm x^{j}(1+r^2)^{-1},\\&&\label{var4}
R=(1+r^2)^{-1}, \quad \theta=\pm\sqrt{2}t, \quad \psi_{j}=\pm x^{j}(1+\xi+r^2)^{-1},\\&&\label{var5}
R=(1+r^2)^{-1}, \quad \theta=\pm(\sqrt{2}+\xi) t, \quad \psi_{j}=\pm x^{j}(1+r^2)^{-1},\\&&\label{var6}
R=(1+r^2)^{-1}+\xi e^{-r^2}, \quad \theta=\pm \sqrt{2} t, \quad \psi_{j}=\pm x^{j}(1+r^2)^{-1},\\&&\label{var7}
R=(1+r^2)^{-1}, \quad \theta=\pm \sqrt{2} t, \quad \psi_{j}=\pm (x^{j}+\xi)(1+r^2)^{-1},\\&&\label{var8}
R=(1+r^2)^{-1}, \quad \theta=\pm \sqrt{2} t, \quad \psi_{j}=\pm x^{j}(1+(r+\xi)^2)^{-1},\\&&\label{var9}
R=(1+(1+\xi)x^2+y^2+z^2)^{-1}, \quad \theta=\pm \sqrt{2} t, \quad \psi_{j}=\pm x^{j}(1+r^2)^{-1},\\&&
R=(1+r^2)^{-1}, \quad \theta=\pm \sqrt{2} t, \quad \psi_{1}=\pm (x+\xi)(1+r^2)^{-1},\nonumber\\&& \label{var10} \quad\quad \psi_{2}=\pm y(1+r^2)^{-1},\quad \psi_{3}=\pm z(1+r^2)^{-1},
\end{eqnarray}
where $\xi$ is a small parameter whose larger values correspond to larger deformations. The case $\xi = 0$ leads to the same special massless  solution (\ref{SxS}). For such arbitrary deformations (\ref{As1}) and (\ref{var1})-(\ref{var10}) at $t=0$,  Fig.~\ref{mnb} demonstrates that a larger deformation leads  to a further  increase  in the total energy, as  expected. Furthermore, it is obvious that
  parameter $B$  has a main role in the stability of  special massless  solution (\ref{SxS}),  and its larger values lead to higher stability (of the special  solution).  To put it differently, the larger the values,  the greater  the increase in the total energy for any arbitrary small variation above the background of  special massless  solution (\ref{SxS}).

Since the theory is relativistic, the moving version of   single massless solitary wave solution (\ref{SxS}) can be easily obtained. Hence, a moving solution along the  $x$-axis is given by    
\begin{eqnarray} \label{fhi}
&& R=\frac{1}{1+\gamma^2 (x-vt)^2+y^2+z^2}, \quad \theta=k_{\mu}x^{\mu}, \quad \psi_{1}=\pm\frac{\gamma (x-vt)}{1+\gamma^2 (x-vt)^2+y^2+z^2},\nonumber\\ \label{fhi2}&&
\psi_{2}=\pm\frac{y}{1+\gamma^2 (x-vt)^2+y^2+z^2},\quad \psi_{3}=\pm\frac{z}{1+\gamma^2 (x-vt)^2+y^2+z^2},
\end{eqnarray}
where $k^{\mu}\equiv (\gamma \omega_{s},\gamma \omega_{s}v,0,0)$.  Since the model  is  relativistic,  the total energy of  moving version (\ref{fhi}) of the special solution (\ref{SxS}) is also zero. In fact, for  moving version  (\ref{fhi}), as well as the static version (\ref{SxS}),  all independent scalars $\mathbb{L}_{i}$ and ${\cal K}_{i}$ ($i=1,\cdots,12$) would be zero simultaneously. Thus,  according to Eqs.~(\ref{eis1})-(\ref{eis12}), the energy density function  and subsequently the total energy, irrespective of the velocity, are   zero.     However, based on all previous knowledge of numerical simulations  about field evolutions in interactions,  we can claim that having a rigid entity without any small  deformation  is generally impossible. In fact,  the internal structure of any solitary wave  solution  would  be slightly  deformed  in the interactions. Therefore,  for  special massless solution (\ref{SxS}), the  rest-mass (energy) is never  absolutely zero. In other words,  the variations of the fields    $\delta R$, $\delta\psi_{j}$ ($j=1,2,3$), and $\delta \theta$ do not remain zero  in the interactions; hence,  $\delta{\cal K}_{i}$, $\delta\varepsilon_{i}$ ($i=1,\cdots,12$), and total energy are not absolute zero. Accordingly,   it is not really a rigid entity with absolute zero rest-mass; therefore,  the effect of any interaction may cause its speed  to approach  the speed of light, but not exactly reach it.

Since  special solution (\ref{SxS}) is non-topological, a multi  particle-like (lump) solution can be easily obtained  only by  adding any arbitrary   number of the  distant (moving) special solutions (\ref{SxS}) together.
In fact,   the non-topological solutions are zero at far distances, hence,  when they are too far apart, the tail of each non-topological solution would be zero in the positions of other solutions. In other words,  the effect of each non-topological solution on the others is practically zero when they are too far apart, similar to many point charges which stand at far distances from one another. For example,  we can consider  two  moving special solutions  (\ref{SxS})   which  initially  stand  at different    positions $(a,0,0)$ and $(b,0,0)$, and have different velocities $\textbf{v}_{1}=-\textbf{v}_{2}=v\widehat{i}$ along the $x$-axis. If $|b-a|$ is large enough,  their linear combination, i.e.
\begin{eqnarray} \label{multis}
	&& R=\frac{1}{1+\gamma^2(x-vt-a)^2+y^2+z^2}+
	\frac{1}{1+\gamma^2(x+vt-b)^2+y^2+z^2}\nonumber
	\\&& \psi_{1}= \frac{\gamma(x-vt-a)}{1+\gamma^2(x-vt-a)^2+y^2+z^2}+
	\frac{\gamma(x+vt-b)}{1+\gamma^2(x+vt-b)^2+y^2+z^2},\nonumber
	\\&&  \psi_{j}= \frac{x^j}{1+\gamma^2(x-vt-a)^2+y^2+z^2}+
	\frac{x^j}{1+\gamma^2(x+vt-b)^2+y^2+z^2}\quad\quad (j=2,3),\nonumber
\end{eqnarray}
is  again a solution  at  the initial times (i.e. the times that are close to $t=0$). For such a linear combination, it is observed numerically that the terms  $\mathbb{L}_{i}$ ($i=1,\cdots,12$) are all approximately zero. Hence, based on dynamical equations (\ref{jkt})-(\ref{jkt2}), such a linear combination would be an approximate   solution again.  The greater the distance between the two special solutions, the more accurate this approximation will be.   It should be noted that for such a linear combination, the velocity-dependent phase-field  $\theta$ changes from  $\theta_{1}=k_{1\mu}x^{\mu}$ ($k_{1\mu}k_{1}^{\mu}=2$) at the position of the first special solution  to  $\theta_{2}=k_{2\mu}x^{\mu}$ ($k_{2\mu}k_{2}^{\mu}=2$) at the position of the second one.  
In fact, in  the free space between two special solutions, where  scalar fields  $R$, $\psi_{j}$ ($j=1,2,3$), and    $\varepsilon$ are all almost zero, there is no rigorous condition on the phase-field to be a solution of   $\mathbb{L}_{2}=0$.


\section{Conclusions}\label{sec5}

For  several  scalar fields $\phi_{i}$ ($i=1,\cdots,N$), we reintroduced the relativistic  $k$-fields systems as  non-standard Lagrangian densities which  are not linear in the kinetic scalars  ${\cal S}_{ij}=\partial_{\mu}\phi_{i}\partial^{\mu}\phi_{j}$. For a group of these systems,  we   showed   that it is possible to have zero rest-mass solutions whose  energy density functions are zero. These massless solutions are not necessarily energetically stable, and finding a stable case is not  simple. Expecting  this stable solution to be a non-topological entity would increase the difficulty of this goal. However,
we introduced a $k$-field system (\ref{kk}) in the $3+1$ dimensions which leads to a single   massless non-topological energetically stable  soliton solution (\ref{SxS}).

Model (\ref{kk}) is based on introducing twelve independent scalar functionals   ${\cal K}_{i}$'s ($i=1,\cdots,12$) of   five scalar fields $R$, $\theta$ and $\psi_{j}$ ($j=1,2,3$).
In general, all terms in the related dynamical equations  (\ref{jkt})-(\ref{jkt2}) contain the first or second power of one of the  ${\cal K}_{i}$'s. Also, all terms in the   energy density function are positive definites and all contain the square of one of the twelve independent functionals ${\cal K}_{i}$'s.
Thus,  the solutions for which all ${\cal K}_{i}$'s  equal  zero simultaneously are  special massless solutions.
Nevertheless, the simultaneous satisfaction of   twelve independent conditions ${\cal K}_{i}=0$  for five scalar fields is not mathematically  possible.
However, we built this model in such a way that  there is  an exceptional      massless solution (\ref{SxS})  for which ${\cal K}_{i}=0$ ($i=1,\cdots,12$).

In general, if there is a rigid massless entity, the effect of any small force changes its speed  to approach the speed of light immediately. 
However, if we assume particles as the soliton  solutions  of the nonlinear field theories, the existence of a rigid particle  would not be possible normally. In other words, they would be deformed   in any interaction, no matter how small. 
Hence,  hypothetical massless particles can never  exactly reach the speed of light. If they exist, they are affected by the environment and their energies would not be absolute zero.



Since  special massless  solution (\ref{SxS}) is single, and since all terms in the energy density function (see Eqs.~(\ref{eis1})-(\ref{eis12})) are positive definites,   the energetical    stability of  special  massless  solution (\ref{SxS}) is guaranteed properly; which means  that, for any arbitrary deformation above the background, the total energy increases.
In other words, the  other solutions of  system (\ref{kexn}) for which    at least one of the ${\cal K}_{i}$'s is a non-zero functional have non-zero positive total energies. Thus, the  energy of the single massless   solution (\ref{SxS}) would be  the least of all solutions.
Accordingly, we can call  special  solution (\ref{SxS}) a (massless) soliton solution.
To summarize, this model shows that  the relativistic classical field theory can lead to   stable  particle-like solutions   with zero rest-masses in $3 + 1$ dimensions.


\section*{Acknowledgement}

The authors  wish  to express their  appreciation to the Persian Gulf University Research Council for their constant support.





\begin{thebibliography}{99}



\bibitem{rajarama} R. Rajaraman \textit{Solitons and instantons} (Amsterdam North Holland,  Elsevier) (1982).





\bibitem{TS}  N. Manton P Sutcliffe \textit{Topological solitons} (Cambridge University Press) (2004).













\bibitem{TA1} A A Izquierdo, J Queiroga-Nunes and L M Nieto P\textit{hysical Review D} \textbf{103} 045003 (2021).

\bibitem{TA2} M Mohammadi and R Dehghani  \textit{Communications in Nonlinear Science and Numerical Simulation} \textbf{94} 105575 (2021).

\bibitem{TA3} D Bazeia, A R Gomes and F C Simas \textit{The European Physical Journal C} \textbf{81} 1 (2021).








































\bibitem{Kink8} V A Gani,  V Lensky and M A Lizunova \textit{Journal of High Energy Physics} \textbf{2015} 147 (2015).










\bibitem{SKrme} T H R Skyrme   \textit{ Proceedings of the Royal Society A.}  \textbf{260} 127 (1961).

\bibitem{SKrme2} T H R Skyrme  \textit{Nuclear Physics}, \textbf{31} 556 (1962).

\bibitem{SKrme3} N S Manton, B J Schroers and M A Singer  \textit{ Communications in mathematical physics} \textbf{245} 123 (2004).







\bibitem{SKrme55} O L Battistel  \textit{Brazilian journal of physics} \textbf{34} 742 (2004).







\bibitem{toft} G 't Hooft  \textit{Nuclear Physics B} \textbf{79} 276 (1974).

\bibitem{pol} A M Polyakov \textit{JETP Letters} \textbf{20} 430 (1974).







\bibitem{TOF} S Nishino, R Matsudo, M Warschinke and K I ondo  \textit{Progress of Theoretical and Experimental Physics} \textbf{2018} 103B04 (2018).

\bibitem{CTOPO}  M Eto, Y Hirono, Nitta and S Yasui  \textit{Progress of Theoretical and Experimental Physics}, \textbf{2014} 012D01 (2014).







\bibitem{Optic1} A R Seadawy and  D Lu \textit{Physical Review E} \textbf{94} 823 (2020).

\bibitem{Optic2} Z Korpinar, M Inc, B Almohsen and M Bayram \textit{Indian Journal of Physics} \textbf{95} 2143 (2021).

\bibitem{Optic3} H Sakaguchi and B A Malomed \textit{Physical Review E}  \textbf{72} 046610 (2005).

\bibitem{Optic4} M Mirzazadeh, M Eslami and A H Arnous \textit{European Physical Journal Plus} \textbf{130} 1 (2015).

\bibitem{Optic5} M Younis, S T R Rizvi \textit{Journal of nanoelectronics and optoelectronics} \textbf{10} 179 (2015).

\bibitem{Optic6} E C Aslan and M Inc \textit{Optik} \textbf{196} 162661 (2019). 


\bibitem{Optic7} W X Ma and M  Chen \textit{Applied Mathematics and Computation}  \textbf{215} 2835 (2009).


\bibitem{Optic8} M Savescu, K R  Khan, P Naruka, H Jafari, L Moraru and A Biswas \textit{Journal of Computational and Theoretical Nanoscience} \textbf{10} 1182 (2013).


\bibitem{Optic9} X Liu, H. Zhang and  W Liu, \textit{Applied Mathematical Modelling}  \textbf{102} 305 (2022).
\bibitem{Optic10} G Ma, J Zhao, Q Zhou, A Biswas and L Liu \textit{Nonlinear Dynamics} \textbf{106} 2479 (2021).


\bibitem{Optic11} L L Wang and W J Liu \textit{Chinese Physics B} \textbf{29} 070502 (2020). 


\bibitem{Optic12} Y Y Yan and W J Liu \textit{Chinese Physics Letters} \textbf{38} 094201 (2021).


\bibitem{Optic13} L Wang, Z Luan, Q Zhou, A Biswas A K Alzahrani and W Liu \textit{Nonlinear dynamics} \textbf{104} 2613 (2021).


\bibitem{Optic14} H Wang, Q Zhou, A Biswas and W Liu \textit{Nonlinear Dynamics} \textbf{106} 841 (2021).


\bibitem{Optic15} G Ma, Q Zhou, W Yu, A Biswas and W Liu \textit{Nonlinear Dynamics} 2509 \textbf{106} (2021).


\bibitem{Optic16} T Y Wang, Q Zhou and W J Liu \textit{Chinese Physics B}  \textbf{31} 020501 (2022).




\bibitem{Am1} A M Wazwaz \textit{Chaos, Solitons $\&$ Fractals} \textbf{37} 1136 (2008).

\bibitem{Am2} Y Zhou, M Wang and T Miao \textit{Physics Letters A} \textbf{323} 77 (2004).

\bibitem{Am3}  N Mahak and G Akram \textit{The European Physical Journal Plus} \textbf{134} 1 (2019).


\bibitem{Am4}   M M El-Borai, H M El-Owaidy, H M Ahmed and A H Arnous \textit{Nonlinear Science Letters A} \textbf{8} 32 (2017).
	



\bibitem{Plasma1} K M Li \textit{Indian Journal of Physics}  \textbf{88} 93 (2014). 
\bibitem{Plasma2} J Akter,  N A Chowdhury,  A Mannan and  A A Mamun  \textit{Indian Journal of Physics}  1 (2021). 

\bibitem{Plasma3} S A El-Tantawy, S A Shan, N Akhtar and A T Elgendy \textit{Chaos, Solitons $\&$ Fractals} \textbf{113} 356 (2018).

\bibitem{Plasma4} M S Ruderman, T Talipova and E Pelinovsky \textit{Journal of Plasma Physics} \textbf{74} 639 (2008).

\bibitem{Plasma5}  P Eslami and M Mottaghizadeh \textit{Indian Journal of Physics} \textbf{88}, 521 (2014).


\bibitem{KDV1}  G Rowlands, P Rozmej, E Infeld and A Karczewska \textit{The European Physical Journal E} \textbf{49} 1 (2017).

\bibitem{KDV2} M K Brun and H Kalisch \textit{Analysis and Mathematical Physics} \textbf{8} 57 (2018).
\bibitem{KDV3}  Z Emami and H R Pakzad \textit{Indian Journal of Physics}, \textbf{85} 1643 (2011).
\bibitem{KDV4}  M Wang \textit{Physics Letters A} \textbf{213} (1996). 

\bibitem{KDV5} H X Ge, R J Cheng and S Q Dai \textit{Physica A: Statistical Mechanics and its Applications} \textbf{357} 466 (2005).








\bibitem{R2}  D Bazeia, L Losano, M A Marques and R Menezes  \textit{Physics Letters B} \textbf{765} 359 (2017).


\bibitem{R8} D Bazeia, L Losano, M A Marques,  R Menezes and R da Rocha  \textit{Physics Letters B} \textbf{758} 146 (2016).



\bibitem{Vak3} A G Panin and M NmSmolyakov  \textit{Physical Review D} \textbf{95} 065006 (2017).

\bibitem{Vak4} A Kovtun, E Nugaev  and  A Shkerin   \textit{Physical Review D} \textbf{98} 096016 (2018).

\bibitem{Vak5} M N Smolyakov  \textit{Physical Review D} \textbf{97} 045011 (2018).

\bibitem{Vak6} M I Tsumagari,  E J Copeland and P M Saffin  \textit{Physical Review D} \textbf{78} 065021 (2008).




























\bibitem{Vak22} M N  Smolyakov \textit{Physical Review D} \textbf{100} 045002 (2019).












\bibitem{Derrick} G H Derrick  \textit{Journal of Mathematical Physics} \textbf{5} 1252 (1964).

\bibitem{MG}  M Mohammadi and R Gheisari \textit{Physica Scripta} \textbf{95} 015301 (2019).


\bibitem{MA} M Mohammadi   \textit{Annals of Physics} \textbf{414} 168099 (2020).

\bibitem{MPS} M Mohammadi  \textit{Physica Scripta} \textbf{95} 045302 (2020).

\bibitem{MPS1} M Mohammadi  \textit{ Annals of Physics} \textbf{422} 168304 (2020).


















\bibitem{Baz1} D Bazeia,  L Losano,  R Menezes and J C R E Oliveira  The European Physical Journal C \textbf{51} 953 (2007).

\bibitem{Adam} C Adam, J Sanchez-Guillen and A Wereszczy\'{n}ski \textit{Journal of Physics A: Mathematical and Theoretical} \textbf{40} 13625 (2007).

\bibitem{Bab} E Babichev \textit{Physical Review D} \textbf{74} 085004 (2006).

























\bibitem{Arnold} V I Arnold, \textit{Mathematical Methods of Classical Mechanics}, Springer, New York, (1978).
\bibitem{DY1} Z E Musielak \textit{Journal of Physics A: Mathematical and Theoretical} \textbf{41} 055205 (2008).
\bibitem{DY2} Z E Musielak, D Roy and  L D Swift \textit{Chaos, Solitons $\&$ Fractals }\textbf{38} 894 (2008).
\bibitem{DY22}  Z E Musielak \textit{Chaos, Solitons  $\&$ Fractals} \textbf{42} 2645 (2009).


\bibitem{DY3}  A R El-Nabulsi \textit{Indian Journal of Physics} \textbf{87} 379 (2013).



\bibitem{DY4} R A El-Nabulsi \textit{Nonlinear Dynamics} \textbf{74} 381 (2013).


\bibitem{DY5} R A El-Nabulsi \textit{Nonlinear Dynamics} \textbf{79} 2055 (2015). 

\bibitem{DY6}   R A El-Nabulsi \textit{Proceedings of the National Academy of Sciences, India Section A: Physical Sciences} \textbf{85} 247 (2015).
\bibitem{DY7} R A El-Nabuls \textit{Qualitative theory of dynamical systems} \textbf{12} 273 (2013).




\bibitem{El1} R A El-Nabulsi \textit{Applied Mathematics Letters} \textbf{43} 120 (2015).

\bibitem{Song} J Song and Y Zhang \textit{Acta Mechanica} \textbf{229} 285 (2018).



\bibitem{Noether} Y Zhou and X S  Zhou \textit{Nonlinear Dynamics} \textbf{84} 1867  (2016).



\bibitem{Geometric0} J F Cari$\tilde{\textrm{n}}$ena and J Fern$\acute{\textrm{a}}$ndez N$\acute{\textrm{u}}$$\tilde{\textrm{n}}$ez \textit{Nonlinear Dynamics} \textbf{83} 457 (2016).

\bibitem{Geometric} J F Cari$\tilde{\textrm{n}}$ena and J  Fern$\acute{\textrm{a}}$ndez N$\acute{\textrm{a}}$$\tilde{\textrm{n}}$ez
\textit{Nonlinear Dynamics} \textbf{86} 1285 (2016).


\bibitem{RAm}  R A El-Nabulsi \textit{Tbilisi} Mathematical Journal \textbf{9} 279 (2016).


\bibitem{Rami} A R El-Nabulsi   \textit{Mathematical Sciences} \textbf{9} 173 (2015).


\bibitem{Zhang} Y Zhang and X P Wang \textit{Symmetry} \textbf{11} 1061 (2019).






\bibitem{Diff1} N A Kudryashov and D I Sinelshchikov \textit{Applied Mathematics Letters} \textbf{63} 124 (2017).

\bibitem{Diff2} J F Cari$\tilde{\textrm{n}}$ena and P Guha \textit{International Journal of Geometric Methods in Modern Physics} \textbf{16} 1940001 (2019).


\bibitem{Diff3} J F Carinena, M F Ranada and M Santander \textit{Journal of mathematical physics} \textbf{46} 062703 (2005).


\bibitem{Diff4} V K Chandrasekar, S N Senthilvelan and M Lakshmanan \textit{Journal of mathematical physics} \textbf{47} 023508 (2006). 


\bibitem{SA2} A Saha and B Talukdar Talukdar \textit{Reports on Mathematical Physics} \textbf{73} 299 (2014).

\bibitem{qu1} R A El-Nabulsi \textit{Indian Journal of Physics} \textbf{87} 379 (2013).
\bibitem{qu2} R A El-Nabulsi \textbf{83} \textit{Proceedings of the National Academy of Sciences, India Section A: Physical Sciences} 383 (2013).
\bibitem{Cfield} R A El-Nabulsi \textit{Indian Journal of Physics} \textbf{87} 465 (2013). 





\bibitem{Cfield2} R A El-Nabulsi Zeitschrift FF$\ddot{\textrm{u}}$r Naturforschung \textbf{71} 817 (2016). 


\bibitem{Armend} C Armend$\acute{\textrm{a}}$riz-Pic$\acute{\textrm{o}}$n, T Damour and  V I Mukhanov \textit{Physics Letters B} \textbf{458} 209 (1999). 

\bibitem{Chiba} T Chiba, T Okabeand  M Yamaguchi  \textit{Physical Review D} \textbf{62} 023511 (2000). 

\bibitem{Armend2} C Armendariz-Picon, V Mukhanov and P J Steinhardt  \textit{Physical Review Letters} \textbf{85} 4438 (2000). 


\bibitem{cosmology1} R A El-Nabulsi  \textit{Journal of the Korean Physical Society}  \textbf{79} 345 (2021). 



\bibitem{Armend3} C Armendariz-Picon and  E A Lim \textit{Journal of Cosmology and Astroparticle Physics }\textbf{2005} 007 (2005). 


\bibitem{cos2} T Padmanabhan and T R Choudhury  \textit{Physical Review D} \textbf{66} 081301 (2002). 
\bibitem{cos3} S Renaux-Petel and G Tasinato  \textit{Journal of Cosmology and Astroparticle Physics} \textbf{2009} 012 (2009). 




\bibitem{Alekseev} A I Alekseev and B A Arbuzov \textit{Theoretical and Mathematical Physics} \textbf{59} 372 (1984). 


\bibitem{Milky} R A El-Nabulsi \textit{Communications in Theoretical Physics} \textbf{69} 233 (2018).

\bibitem{ple} R A El-Nabulsi \textit{Proceedings of the Royal Society A} \textbf{476} 20200190 (2020).

\bibitem{cos4} R A El-Nabulsi  \textit{Journal of Theoretical and Applied Physics }\textbf{7} 58 (2013).











\bibitem{Rami2}  A R El-Nabulsi \textit{Canadian Journal of Physics }\textbf{92} 1149 (2014).



\end{thebibliography}
\end{document}